\documentclass[12pt]{article}

\usepackage{graphicx}
\usepackage{amssymb}
\begin{document}

\newcommand{\vecx}{\mbox{\boldmath $x$}}


%
\begin{center}
{\large\bf
Bose-Einstein and Fermi-Dirac distributions in nonextensive quantum 
statistics: Exact and interpolation approaches
} 
\end{center}

\begin{center}
Hideo Hasegawa
\footnote{hideohasegawa@goo.jp}
\end{center}

\begin{center}
{\it Department of Physics, Tokyo Gakugei University,  \\
Koganei, Tokyo 184-8501, Japan}
\end{center}
\begin{center}
({\today})
\end{center}
\thispagestyle{myheadings}

\begin{abstract}
Generalized Bose-Einstein (BE) and Fermi-Dirac (FD) distributions 
in nonextensive quantum statistics have been discussed  
by the maximum-entropy method (MEM) with the optimum Lagrange multiplier
based on the exact integral representation 
[Rajagopal, Mendes, and Lenzi, Phys. Rev. Lett. {\bf 80}, 3907 (1998)].
It has been shown that the $(q-1)$ expansion in the exact approach
agrees with the result obtained by the asymptotic approach
valid for $O(q-1)$.
Model calculations have been made 
with a uniform density of states for electrons and 
with the Debye model for phonons.
Based on the result of the exact approach, 
we have proposed the {\it interpolation
approximation} to the generalized distributions, which
yields results in agreement with the exact approach within 
$O(q-1)$ and in high- and low-temperature limits.
By using the four methods of the exact,
interpolation, factorization and superstatistical approaches,
we have calculated coefficients in the generalized Sommerfeld expansion,
and electronic and phonon specific heats at low temperatures.
A comparison among the four methods  
has shown that the interpolation approximation is potentially 
useful in the nonextensive quantum statistics.
Supplementary discussions have been made on the $(q-1)$ expansion of 
the generalized distributions based on the exact approach
with the use of the un-normalized MEM, 
whose results also agree with those of the asymptotic approach.

\end{abstract}

\vspace{0.5cm}
\vspace{1cm}

 

\newpage

\section{INTRODUCTION}

In the last decade, many studies have been made
for the nonextensive statistics \cite{Tsallis88} in which 
the generalized entropy (the Tsallis entropy) is introduced 
(for a recent review, see \cite{Tsallis04}).
The Tsallis entropy is a one-parameter generalization of the Boltzmann-Gibbs 
entropy with the entropic index $q$: the Tsallis entropy
in the limit of $q=1.0$ reduces to the Boltzmann-Gibbs entropy.
The optimum probability distribution or density matrix
is obtained with the maximum entropy method (MEM) for the Tsallis entropy
with some constraints. At the moment, there are four possible 
MEMs: original method \cite{Tsallis88},
un-normalized method \cite{Curado91}, normalized method \cite{Tsallis98}, 
and the optimal Lagrange multiplier (OLM) method \cite{Martinez00}.
The four methods are equivalent in the sense that distributions 
derived in them are easily transformed each other \cite{Ferri05}.
A comparison among the four MEMs is made in Ref. \cite{Tsallis04}.
The nonextensive statistics has been successfully applied 
to a wide class of subjects in physics, chemistry, information science,
biology and economics \cite{NES}.

One of alternative approaches to the nonextensive statistics 
besides the MEM is the superstatistics \cite{Wilk00,Beck01}
(for a recent review, see \cite{Beck07}).
In the superstatistics, it is assumed that locally 
the equilibrium state is described by the Boltzmann-Gibbs statistics 
and that their global properties may be expressed by a superposition 
over the intensive parameter ({\it i.e.,} the inverse temperature) 
\cite{Wilk00}-\cite{Beck07}.
It is, however, not clear how to obtain the mixing probability 
distribution of fluctuating parameter from first principles.
This problem is currently controversial and some attempts to 
this direction have been proposed \cite{Souza03}-\cite{Straeten08}.
The concept of the superstatistics has been applied to many kinds 
of subjects such as hydrodynamic turbulence \cite{Beck03,Reynolds03,Beck07b}, 
cosmic ray \cite{Beck04} and solar flares \cite{Baies06}.

The nonextensive statistics has been applied to both classical and quantum systems. 
In this paper, we pay attention to quantum nonextensive systems.
The generalized Bose-Einstein and Fermi-Dirac distributions
in nonextensive systems (referred to as $q$-BED and $q$-FDD 
hereafter) have been discussed by the three methods.
(i) The asymptotic approximation (AA) was proposed by
Tsallis, Sa Barreto and Loh \cite{Tsallis95} who derived
the expression for the canonical partition function valid for 
$ \vert q-1 \vert/k_B T \rightarrow 0$.
It has been applied to the black-body radiation \cite{Tsallis95},
early universe \cite{Tsallis95,Torres97} and
the Bose-Einstein condensation \cite{Tsallis95}\cite{Tirnakli00}.
(ii) The factorization approximation (FA) was proposed 
by B\"{u}y\"{u}kkilic, Demirhan and G\"{u}lec \cite{Buy95}
to evaluate the grand canonical partition function.
The FA was criticized in \cite{Pennini95}\cite{Lenzi01}, but 
supported in \cite{Wang97},
related discussion being given in Sec. 4.
The simple expressions for $q$-BED and $q$-FDD 
in the FA have been adopted in many applications
such as the black-body radiation 
\cite{Tirnakli00,Tirnakli97,Wang98,Buy02},
early universe \cite{Tirnakli99,Pessah01},
the Bose-Einstein condensation \cite{Torres98}-\cite{Lawani08},
metals \cite{Oliveira00}, superconductivity \cite{Nunes01,Uys01},
spin systems \cite{Portesi95}-\cite{Reis06} and
metallic ferromagnets \cite{Hasegawa09}.
(iii) The exact approach (EA) was developed by Rajagopal, Mendes and Lenzi
\cite{Rajagopal98,Lenzi99} who derived the formally exact
integral representation for the grand canonical partition
function of nonextensive systems which is 
expressed in terms of the Boltzmann-Gibbs counterpart.
The integral representation approach originated 
from the Hilhorst formula \cite{Prato95}.
Because an actual evaluation of a given integral is generally
difficult, it may be performed in an approximate way 
\cite{Rajagopal98,Lenzi99} or in the limited cases \cite{Rego03}.
The validity of the EA is discussed in \cite{Solis03,Martinez03}.
The EA has been applied to 
nonextensive quantum systems such as 
black body radiation \cite{Lenzi98,Martinez02}
and the Bose-Einstein condensation \cite{Rajagopal98,Lenzi99}. 

We believe that it is important and valuable to pursue the EA 
despite its difficulty. It is the purpose of the present study 
to apply the EA \cite{Rajagopal98,Lenzi99} to
calculations of the generalized distributions of $q$-BED and $q$-FDD. 
The grand canonical partition function of the nonextensive systems is derived
with the use of the OLM scheme in the MEM \cite{Martinez00}.
Self-consistent equations for averages of the number of particles and energy and 
the grand-canonical partition function are exactly expressed by 
the integral representation \cite{Rajagopal98,Lenzi99}.
The integral representation for $q > 1.0$ in the EA
is expressed as an integral along the real axis, while
that for $q < 1.0$ 
is expressed as the contour integral in the complex plane
\cite{Rajagopal98,Lenzi99,Rego03}.
We have shown that  the $(q-1)$ expansion by the EA agrees  
with the result derived by the AA.
For $q \geq 1.0$, the self-consistent equations
have been numerically solved with the band model 
for electrons and the Debye model for phonon. 

It is rather difficult and tedious to obtain the generalized distributions 
in the EA because they need the self-consistent calculation of averages of
number of particles and energy.
Based on the exact result obtained, we have proposed
the {\it interpolation approximation} (IA) to $q$-BED and $q$-FDD,
which do not need the self-consistently determined quantities and
whose results are in agreement with those of the EA
within $O(q-1)$ and in high- and low-temperature limits.
We may obtain the simple analytic expressions of the $q$-BED and $q$-FDD.

The paper is organized as follows.
In Sec. 2, the exact integral representation is derived 
with the OLM-MEM after
Ref. \cite{Rajagopal98,Lenzi99,Rego03}.
We have discussed the $(q-1)$ expansion of physical quantities,
using the EA and AA. 
Numerical calculations are performed for electron and phonon models,
for which we present the $q$-BED and $q$-FDD with the temperature-dependent energy. 
In Sec. 3, we propose the IA, by which 
analytical expressions for $q$-BED and $q$-FDD are obtained.
In Sec. 4, a comparison is made between the generalized distributions 
calculated by the four methods of the EA, IA, FA \cite{Buy95} 
and the superstatistical approximation (SA).
A controversy on the validity of the FA \cite{Buy95} is discussed.
With the use of the four methods,
the generalized Sommerfeld expansion, and low-temperature electronic
and phonon specific heats are calculated.
Sec. 5 is devoted to our conclusion. 
In Appendix A, we present a study 
of the EA and AA with the un-normalized
MEM \cite{Curado91,Tsallis95}, calculating the $(q-1)$ 
expansion of the $q$-BED and $q$-FDD.
Supplementary discussions on the IA are presented in Appendix B.


\section{EXACT APPROACH}
\subsection{MEM by OLM}

We will study nonextensive quantum systems described by the
hamiltonian $\hat{H}$.
We have obtained the optimum density matrix of $\hat{\rho}$, 
applying the OLM-MEM to the Tsallis entropy 
given by 
\cite{Martinez00,Ferri05}
\begin{eqnarray}
S_q &=& \frac{k_B}{q-1}[1- Tr \hat{\rho}_q^q], \nonumber
\end{eqnarray}
with the constraints:
\begin{eqnarray}
Tr \hat{\rho}_q &=& 1, \nonumber \\
Tr \{ \hat{\rho}_q^q N \} &=& c_q N_q, \nonumber  \\
Tr \{ \hat{\rho}_q^q H \} &=& c_q E_q, \nonumber  \\
c_q &=& Tr \hat{\rho}_q^q, \nonumber 
\end{eqnarray}
where $Tr$ stands for the trace, 
$k_B$ is the Boltzmann constant, and
$E_q$ and $N_q$ denote the expectation values
of the hamiltonian $\hat{H}$ and the number operator $\hat{N}$,
respectively.
The OLM-MEM yields \cite{Martinez00,Ferri05}
\begin{eqnarray}
\hat{\rho}_q &=& \frac{1}{X_q} 
[1+(q-1) \beta(\hat{H}-\mu \hat{N} - E_q + \mu N_q)]^{\frac{1}{1-q}}, 
\label{eq:A3}\\
X_q &=& 
Tr \{ [1+(q-1) \beta(\hat{H}-\mu \hat{N} - E_q + \mu N_q)]^{\frac{1}{1-q}} \}, 
\label{eq:A4}\\
N_q &=& \frac{1}{X_q} 
Tr \{ [1+(q-1) \beta(\hat{H}-\mu \hat{N} - E_q + \mu N_q)]^{\frac{q}{1-q}}\:N \}, 
\label{eq:A5}\\
E_q &=& \frac{1}{X_q} 
Tr \{[1+(q-1) \beta(\hat{H}-\mu \hat{N} - E_q + \mu N_q)]^{\frac{q}{1-q}} \:H \},
\label{eq:A6}
\end{eqnarray}
where $\beta$ and $\mu$ denote the Lagrange multipliers.
In deriving Eqs. (\ref{eq:A3})-(\ref{eq:A6}), we have employed the relation:
\begin{eqnarray}
c_q &=& X_q^{1-q}.
\label{eq:A9} \nonumber 
\end{eqnarray}
Lagrange multipliers of $\beta$ and $\mu$ are identified
as the inverse physical temperature ($\beta=1/k_B T$)
and the chemical potential (fermi level), respectively.
\cite{Martinez00,Ferri05}.

\subsection{Exact integral representation}

\subsubsection{Case of $q > 1$}

In the case of $q > 1.0$, we adopt the formula for 
the gamma function $\Gamma(s)$:
\begin{eqnarray}
x^{-s} &=& \frac{1}{\Gamma(s)} 
\int_0^{\infty} u^{s-1} e^{-x u} \:du 
\hspace{1cm} \mbox{for $ \Re \:s > 0$}.
\label{eq:B1}
\end{eqnarray}
With $s=1/(q-1)$ [or $s=q/(q-1)$] and 
$x=1+(q-1) \beta (H-\mu N)$ in Eq. (\ref{eq:B1}), 
we may express Eqs. (\ref{eq:A3})-(\ref{eq:A6}) by \cite{Rajagopal98,Lenzi99}
\begin{eqnarray}
N_q &=& \frac{1}{X_q} \int_0^{\infty} 
G\left(u;\frac{q}{q-1},1 \right)
e^{(q-1)\beta u (E_q-\mu N_q)}  \Xi_1[(q-1) \beta u] 
N_1[(q-1) \beta u]\:du, \nonumber \\
&& \label{eq:B2} \\
E_q &=& \frac{1}{X_q} \int_0^{\infty} 
G\left(u;\frac{q}{q-1},1 \right)
e^{(q-1)\beta u (E_q-\mu N_q)}  \Xi_1[(q-1) \beta u] 
E_1[(q-1) \beta u]\:du, \nonumber \\
&& \label{eq:B3}
\end{eqnarray}
with
\begin{eqnarray}
X_q &=& \int_0^{\infty} G\left(u;\frac{1}{q-1},1 \right)
e^{(q-1)\beta u(E_q-\mu N_q)} 
\; \Xi_1[(q-1) \beta u] \:du,
\label{eq:B4}
\end{eqnarray}
where
\begin{eqnarray}
\Xi_1(u) &=& e^{- u \:\Omega_1(u)}
=Tr \{e^{-u (\hat{H}-\mu \hat{N})} \}
= \prod_k [1 \mp e^{-u(\epsilon_k-\mu)}]^{\mp 1},
\label{eq:B5}\\
\Omega_1(u) &=& \pm \frac{1}{u}
\sum_k \ln[1 \mp e^{-u(\epsilon_k-\mu)}], 
\label{eq:B6}\\
N_1(u) &=& \sum_k f_1(\epsilon_k,u), 
\label{eq:B8}\\
E_1(u) &=& \sum_k \epsilon_k f_1(\epsilon_k,u), 
\label{eq:B9}\\
f_1(\epsilon,u) &=& \frac{1}{e^{u(\epsilon-\mu)}\mp 1},
\label{eq:B10} \\
G\left(u;a,b \right) 
&=& \frac{b^a}{\Gamma\left(a  \right)} 
u^{a-1} e^{-b u}.
\label{eq:B7}
\end{eqnarray}
The upper (lower) sign in Eqs. (\ref{eq:B5}), (\ref{eq:B6}) and (\ref{eq:B10}) 
denotes boson (fermion) case,
and $\Xi_1(u)$, $\Omega_1(u)$, $N_1(u)$, $E_1(u)$ and $f_1(\epsilon,u)$ 
express the physical quantities for $q=1.0$.
Equations (\ref{eq:B2})-(\ref{eq:B4}) show that physical quantities
in nonextensive systems
are expressed as a superposition of those for $q=1.0$.

Although Eqs. (\ref{eq:B2})-(\ref{eq:B4}) are formally exact
expressions, they have a problem when we perform numerical calculations. 
The gamma distribution of $G[u;1/(q-1)+\ell, 1]$ ($\ell=0,1$)
in Eqs. (\ref{eq:B2})-(\ref{eq:B4}) has the maximum at $u_{max}$, 
and average and variance given by
\begin{eqnarray}
u_{max} &=& \frac{1}{(q-1)}+\ell-1, 
\label{eq:B11} \\
\langle u  \rangle_u &=&
\frac{1}{(q-1)}+\ell, 
\label{eq:B12} \\
\langle u^2  \rangle_u - \langle u  \rangle_u^2
&=& \frac{1}{(q-1)}+\ell.
\label{eq:B13}
\end{eqnarray}
Equation (\ref{eq:B11}) shows that the gamma distribution 
in Eqs. (\ref{eq:B2})-(\ref{eq:B4}) has the maximum at 
$u_{max} = 1/(q-1) \rightarrow \infty$
while the contribution from $\Xi_1[(q-1)\beta t]$
is dominant at $t \sim 0$ because
its argument becomes $(q-1) \beta t \rightarrow 0$. 
Then numerical calculations using Eqs. (\ref{eq:B2})-(\ref{eq:B4})
are very difficult.

In order to overcome this difficulty,
we have adopted a change of variable: $(q-1) \beta u \rightarrow u $ 
in Eq. (\ref{eq:B2})-(\ref{eq:B4})
to obtain alternative expressions given  by
\begin{eqnarray}
N_q &=& \frac{1}{X_q} \int_0^{\infty} 
G\left(u;\frac{1}{q-1}+1, \frac{1}{(q-1)\beta} \right) \:
e^{u(E_q-\mu N_q)} \: \Xi_1(u) N_1(u) \:du, 
\label{eq:B14} \\
E_q &=& \frac{1}{X_q} \int_0^{\infty} 
G\left(u;\frac{1}{q-1}+1, \frac{1}{(q-1)\beta} \right) \:
e^{u(E_q-\mu N_q)} \: \Xi_1(u) E_1(u) \:du, 
\label{eq:B15}
\end{eqnarray}
with 
\begin{eqnarray}
X_q &=& \int_0^{\infty} 
G\left(u;\frac{1}{q-1}, \frac{1}{(q-1)\beta} \right) \: 
e^{u(E_q-\mu N_q)} \:\Xi_1(u) \:du.
\label{eq:B16} 
\end{eqnarray}
The gamma distribution of $G(u;\frac{1}{(q-1)}+\ell,\frac{1}{q-)} \beta)$ 
for $\ell=0,1$ in Eqs. (\ref{eq:B14})-(\ref{eq:B16}) has the maximum at
$u_{max}$, and 
average, mean square and variance given by 
\begin{eqnarray}
u_{max} &=& [1+ (q-1)(\ell-1)] \beta, 
\label{eq:B17}\\
\langle u \rangle_u &=& [1+(q-1) \ell]\beta,
\label{eq:B18} \\
\langle u^2 \rangle_u &=& [1+(q-1) \ell][1+(q-1)(\ell+1)]\beta^2,
\label{eq:B19}\\
\langle u^2 \rangle_u -\langle u \rangle_u^2
&=& (q-1)[1+(q-1)\ell] \beta^2.
\end{eqnarray} 
Equation (\ref{eq:B17}) shows that
the gamma distribution has the maximum at $u_{max} = \beta$ in the limit
of $q \rightarrow 1.0$,
and an integration over $u$ in Eqs. (\ref{eq:B14})-(\ref{eq:B16})
may be easily performed.
Indeed, in the case of $q \gtrsim 1.0$ discussed above,
the gamma distribution in Eqs. (\ref{eq:B14})-(\ref{eq:B16}) 
becomes
\begin{eqnarray}
G\left( u; \frac{1}{q-1}+\ell, \frac{1}{(q-1)\beta} \right) 
&\rightarrow& \frac{1}{\sqrt{2 \pi (q-1) \beta^2}}
e^{-\frac{1}{2(q-1) \beta^2}\:(u-\beta)^2}, \\
&\rightarrow& \delta(u-\beta) 
\hspace{1cm} \mbox{for $(q-1)\beta^2 \rightarrow 0$}. 
\label{eq:B20}
\end{eqnarray}
Although expressions given by Eqs. (\ref{eq:B2})-(\ref{eq:B4}) are mathematically
equivalent to those given by Eqs. (\ref{eq:B14})-(\ref{eq:B16}),
the latter expressions are more suitable than 
the former ones for numerical calculations. 

\subsubsection{Case of $q < 1$}
In the case of $q < 1.0$, we adopt the formula given by 
\begin{eqnarray}
x^{s} &=& \frac{i}{2 \pi} \Gamma(s+1)\int_C (-t)^{-s-1} e^{-xt}\:dt
\hspace{1cm} \mbox{for $ \Re \:s > 0$},
\label{eq:H1}
\end{eqnarray}
where a contour integral is performed over the Hankel path $C$ 
in the complex plane. 
With $s=1/(1-q)$ [or $s=q/(1-q)$] and 
$x=1+(q-1) \beta (H-\mu N)$ in Eq. (\ref{eq:H1}), 
we obtain \cite{Rajagopal98,Lenzi99}
\begin{eqnarray}
N_q &=& \frac{i}{2 \pi X_q} 
\int_C H\left(t;\frac{q}{1-q},1 \right) 
e^{-(1-q)\beta t (E_q-\mu N_q)}
\Xi_1[-(1-q) \beta t] \:N_1[-(1-q) \beta t]\: dt, \nonumber \\
&& 
\label{eq:H2}\\
E_q &=& \frac{i}{2 \pi X_q} 
\int_C  H\left(t;\frac{q}{1-q},1 \right) 
e^{-(1-q)\beta t (E_q-\mu N_q)}
\Xi_1[-(1-q) \beta t] \:E_1[-(1-q) \beta t]\: dt, \nonumber \\
&& 
\label{eq:H3} 
\end{eqnarray}
with
\begin{eqnarray}
X_q &=& \frac{i}{2 \pi } \int_C H\left(t;\frac{1}{1-q},1 \right)
e^{-(1-q)\beta t(E_q-\mu N_q)} 
\; \Xi_1[-(1-q) \beta t] \:dt, 
\label{eq:H5}\\
H(t;a,b)&=&  \: \Gamma(a+1) b^{-a} \:(-t)^{-a-1} e^{- b t},
\label{eq:H6}
\end{eqnarray}
where $\Xi_1(u)$, $N_1(u)$ $E_1(u)$ and $f_1(\epsilon, u)$ 
are given by  Eqs. (\ref{eq:B5})-(\ref{eq:B9}) with complex $u$.

In the case of $q < 1.0$, $N_q$, $E_q$ and $X_q$ 
given by Eqs. (\ref{eq:H2})-(\ref{eq:H5}) 
are expressed by an integral along the Hankel contour path $C$
in the complex plane.
The Hankel path may be modified to the Bromwich contour which is
parallel to the imaginary axis from $c-i \:\infty$ to $c + i\: \infty$
($c > 0$) \cite{Lenzi98,Martinez02}.
The Bromwich contour is usually understood as counting the
contributions from the residues of all poles located in the left-side
of $ \Re \;z < c$ of the complex plane $z$,
when the integrand is expressed by simple analytic functions.
If the integrand is not expressed by simple analytic functions,
we have to evaluate it by numerical methods. Unfortunately,
we have not succeeded in evaluating Eqs. (\ref{eq:H2})-(\ref{eq:H5}) 
with the sufficient accuracy.
It is not easy to numerically evaluate the integral 
along the Hankel or Bromwich contour, 
which is required to be appropriately deformed 
for actual numerical calculations \cite{Laplace,Schmelzer07}.
This subject has a long history and it is still active in the field 
of the numerical methods
for the inverse Laplace transformation \cite{Laplace} and
for the Gamma functions \cite{Schmelzer07}. 

It is worthwhile to remark that
for a bose gas model with the density of states of
$\rho(\epsilon)=A \epsilon^r$, we obtain (with $\mu=0$)
\cite{Prato95,Lenzi98,Martinez02}
\begin{eqnarray}
\Xi_1(u) &=& \exp\left[\frac{A \Gamma(r+1)\zeta(r+2)}
{u^{r+1}} \right], \nonumber \\
N_1(u) &=& \frac{A \Gamma(r+1)\zeta(r+1)}{u^{r+1}}, \nonumber \\
E_1(u) &=& \frac{A \Gamma(r+2)\zeta(r+2)}{u^{r+2}}, \nonumber
\end{eqnarray}
where $r=1/2$ for an ideal bose gas, $r=2$ for a harmonic oscillator,
$A$ denotes a relevant factor and $\zeta(z)$
stands for the Riemann zeta function. 
With a repeated use of Eq. (\ref{eq:H1}),
$N_q$, $E_q$ and $X_q$ may be expressed as sums of gamma functions
\cite{Prato95,Lenzi98,Martinez02}. 
Unfortunately, such a sophisticated method cannot be necessarily 
applied to any models like a fermi gas. 

With a change of variable of $(1-q)\beta (-t) \rightarrow (-t)$ 
in Eqs. (\ref{eq:H2})-(\ref{eq:H5}) after the case of $q>1$, 
they are given by
\begin{eqnarray}
N_q &=& \frac{i}{2 \pi X_q} 
\int_C H\left(t;\frac{1}{1-q}-1,\frac{1}{(1-q)\beta} \right) 
e^{-t (E_q-\mu N_q)}
\Xi_1(-t) \:N_1(-t)\: dt, \nonumber \\
&& 
\label{eq:H12}\\
E_q &=& \frac{i}{2 \pi X_q} 
\int_C  H\left(t;\frac{1}{1-q}-1,\frac{1}{(1-q)\beta} \right) 
e^{-t (E_q-\mu N_q)} \Xi_1(-t) \:E_1(-t)\: dt, \nonumber \\
&& 
\label{eq:H13} 
\end{eqnarray}
with
\begin{eqnarray}
X_q &=& \frac{i}{2 \pi } 
\int_C H\left(t;\frac{1}{1-q},\frac{1}{(1-q)\beta} \right)
e^{-t(E_q-\mu N_q)} \; \Xi_1(-t) \:dt. 
\label{eq:H15}
\end{eqnarray}
Average and mean square over $H(t,\frac{1}{1-q}-\ell,\frac{1}{(1-q)\beta})$
for $\ell=0,1$ are given by
\begin{eqnarray}
\langle (-t) \rangle_t&=& [1-(1-q) \ell] \beta, 
\label{eq:T7} \\
\langle (-t)^2 \rangle_t&=& [1-(1-q) \ell][q-(1-q)\ell] \beta^2.
\label{eq:T8}
\end{eqnarray}
Equations (\ref{eq:H12})-(\ref{eq:H15}) are useful in making 
the $(q-1)$ expansion, as will be discussed in the following.

\subsection{The $(q-1)$ expansion}
\subsubsection{The exact approach}

We will consider the $(q-1)$ expansion of the expectation value 
of an operator $\hat{O}$ in the EA. By using Eqs. (\ref{eq:B14}) 
and (\ref{eq:H12}), we obtain
\begin{eqnarray}
\langle \hat{O} \rangle_q 
&=& \frac{1}{X_q}
\:Tr\: \{[1-(1-q) \beta \hat{K}]^{\frac{q}{1-q}}\:\hat{O} \},
\label{eq:W2}\\
&=& \frac{1}{X_q} \int_0^{\infty}
G\left( u;\frac{q}{q-1},\frac{1}{(q-1)\beta} \right) Y_1(u) \:O_1(u)\:du
\hspace{1cm}\mbox{for $q > 1$}, \\
&=& \frac{i}{2 \pi X_q} 
\int_C H\left(t;\frac{q}{1-q}, \frac{1}{(1-q)\beta} \right)
Y_1(-t) \:O_1(-t) \:dt
\hspace{0.5cm}\mbox{for $q < 1$},
\end{eqnarray}
with
\begin{eqnarray}
O_1(u) &=& \frac{Tr \{e^{-u \hat{K}} \:\hat{O} \} }{Y_1(u)}, \\
Y_1(u) &=&  Tr \{e^{-u \hat{K}} \}
=e^{u(E_q-\mu N_q)} \:\Xi_1(u), \\
\hat{K} &=& \hat{H}- \mu \hat{N}-E_q+\mu N_q, 
\end{eqnarray}
where $X_q$ is given by Eq. (\ref{eq:B16}) for $q>1$ and
by Eq. (\ref{eq:H15}) for $q < 1$. 
It is noted that $Y_1(u)$ includes the self-consistently calculated
$N_q$ and $E_q$.

We first consider the case of $q \gtrsim 1$ for which the integral including 
an arbitrary function $W(u)$ is assumed to be given by
\begin{eqnarray}
J &=& \int_0^{\infty} G\left(u;\frac{1}{q-1}+\ell,\frac{1}{(q-1)\beta} \right)
W(u) \: du
\hspace{1cm}\mbox{for $\ell=0,1$}.
\label{eq:T1}
\end{eqnarray}
Since $G(u; \frac{1}{q-1}+\ell,\frac{1}{(q-1)\beta})$ has
the maximum around $u=\beta$ as mentioned before [Eq. (\ref{eq:B17})], 
$W(u)$ may be expanded as
\begin{eqnarray}
W(u) 
&=& W(\beta)+(u-\beta)\frac{\partial W}{\partial \beta}
+ \frac{1}{2}(u-\beta)^2 \frac{\partial^2 W}{\partial \beta^2}+\cdot\cdot.
\label{eq:T2}
\end{eqnarray}
Substituting Eq. (\ref{eq:T2}) to Eq. (\ref{eq:T1}) and
using the relations given by Eqs. (\ref{eq:B18}) and (\ref{eq:B19}), 
we  obtain $J$ in a series of $(q-1)$ as
\begin{eqnarray}
J 
&=& W(\beta)+ \langle (u-\beta) \rangle_u \frac{\partial W}{\partial \beta}
+ \frac{1}{2} \langle (u-\beta)^2 \rangle_u \frac{\partial^2 W}{\partial \beta^2} 
+\cdot\cdot,  \\
&=& W(\beta)+ (q-1)\left[\ell \beta \frac{\partial W}{\partial \beta}
+ \frac{1}{2} \beta^2  \frac{\partial^2 W}{\partial \beta^2} \right]
+\cdot\cdot.  
\hspace{1cm}\mbox{for $q \simeq 1.0$}.
\label{eq:T3}
\end{eqnarray} 

Next we consider the case of $q \lesssim 1$ for which a similar integral 
along the Hankel path C is given by
\begin{eqnarray}
J &=& \frac{i}{2 \pi} 
\int_C H \left(t; \frac{q}{1-q}-\ell,\frac{1}{(1-q)\beta} \right) W(-t) \:dt
\hspace{1cm}\mbox{for $\ell=0,1$}.
\label{eq:T4}
\end{eqnarray}
By expanding $W(-t)$ at $-t=\beta$
and using the relations for averages given by Eqs. (\ref{eq:T7}) and (\ref{eq:T8}),
we obtain the same expression for $J$ as Eq. (\ref{eq:T3}),
which is then valid both for $q \lesssim 1.0$ and $ q \gtrsim 1.0$.

For $W(u)= Y_1(u)$ 
and $W(u)= Y_1(u) O_1(u)$ in Eq. (\ref{eq:T3}), 
we obtain 
\begin{eqnarray}
X_q &=& Y_1 + \frac{1}{2}(q-1) \beta^2 \frac{\partial^2 Y_1}{\partial \beta^2}
+ \cdot\cdot, 
\label{eq:T20}
\\
O_q &=& \frac{1}{X_q} \left[Y_1 O_1
+ (q-1) \beta \frac{\partial (Y_1 O_1)}{\partial \beta}
+ \frac{1}{2}(q-1) \beta^2 \frac{\partial^2 (Y_1 O_1)}{\partial \beta^2}
+\cdot\cdot \right].
\label{eq:T21} 
\end{eqnarray}
Note that the $O((q-1)\beta)$ term in Eq. (\ref{eq:T20}) vanishes 
because $\ell=0$ in Eq. (\ref{eq:T3}). 
Substituting the relations given by
\begin{eqnarray}
\frac{\partial Y_1}{\partial \beta} 
&=& - \langle \hat{K} \rangle_1 Y_1,\\
\frac{\partial^2 Y_1}{\partial \beta^2} 
&=& \langle \hat{K}^2 \rangle_1 Y_1,\\
\frac{\partial O_1}{\partial \beta} 
&=& \langle \hat{K} \rangle_1 \langle \hat{O} \rangle_1
- \langle \hat{K}\hat{O} \rangle_1,\\
\frac{\partial^2 O_1}{\partial \beta^2} 
&=& \langle \hat{K}^2\hat{O} \rangle_1 
- \langle \hat{K}^2 \rangle_1 \langle \hat{O} \rangle_1
+ 2[\langle \hat{K} \rangle_1 \langle \hat{K}\hat{O} \rangle_1
- \langle \hat{K} \rangle_1^2 \langle \hat{O} \rangle_1],
\end{eqnarray} 
to Eqs. (\ref{eq:T20}) and (\ref{eq:T21}), we finally obtain
the $O(q-1)$ expansion of $O_q$ given by
\begin{eqnarray}
O_q &\simeq& O_1 + (1-q) \left( \beta \langle \hat{K}\hat{O} \rangle_1
+\frac{1}{2} \beta^2 
[\langle \hat{K}^2 \rangle_1 \langle \hat{O} \rangle_1
-\langle \hat{K}^2 \hat{O} \rangle_1 ] \right) +\cdot\cdot.
\label{eq:T22}
\end{eqnarray}

\subsubsection{The asymptotic approach}
On the other hand, we may adopt the AA \cite{Tsallis95}
to obtain $O_q$ given by Eq. (\ref{eq:W2}) valid for $O(q-1)$.
By using the relation: $e_q^x \simeq e^x [1- (1-q) x^2/2+\cdot\cdot] $
in Eqs. (\ref{eq:A4}) and (\ref{eq:W2}), we may expand $X_q$ and $O_q$ 
up to $O(q-1)$ as
\begin{eqnarray}
X_q &\simeq& X_1 \left[1-\frac{1}{2} (1-q) \beta^2 \langle \hat{K}^2 \rangle_1 
+ \cdot\cdot \right], 
\label{eq:W4} \\
O_q &=& \frac{1}{X_q} 
Tr \{ [1-(1-q) \beta \hat{K}]^{-1}
\:[1-(1-q) \beta \hat{K}]^{\frac{1}{1-q}} \:\hat{O} \},
\label{eq:W5} \\
&\simeq& \frac{1}{X_q}
Tr \{ e^{-\beta \hat{K}}[1+(1-q) \beta \hat{K}]
\left[1 -\frac{1}{2}(1-q)\beta^2 \hat{K}^2 \right] \hat{O} \}+\cdot\cdot, \\
&\simeq& O_1 + (1-q) \left( \beta \langle \hat{K}\hat{O} \rangle_1
+\frac{1}{2} \beta^2 
\left[ \langle \hat{K}^2 \rangle_1 \langle \hat{O} \rangle_1
-\langle \hat{K}^2 \hat{O} \rangle_1 \right] \right)+\cdot\cdot.
\label{eq:W6}
\end{eqnarray}
Equation (\ref{eq:W6}) agrees with Eq. (\ref{eq:T22}) 
obtained by the EA within $O(q-1)$.
In Appendix A, we have shown that the same equivalence holds
between the AA and EA with the un-normalized MEM \cite{Curado91,Tsallis95}.

\subsection{Generalized distributions}
\subsubsection{The $O(q-1)$ expansion}
Equations for $N_q$ and $E_q$ given by Eqs. (\ref{eq:B14}), (\ref{eq:B15}), 
(\ref{eq:H12}) and (\ref{eq:H13}) may be expressed as
\begin{eqnarray}
N_q &=& \sum_k f_q(\epsilon_k, \beta)
= \int \: f_q(\epsilon, \beta) \rho(\epsilon)\: d\epsilon, 
\label{eq:C1} \\
E_q &=& \sum_k f_q(\epsilon_k, \beta) \:\epsilon_k
= \int \: f_q(\epsilon, \beta)\:\epsilon \rho(\epsilon)\: d\epsilon, 
\label{eq:C2}
\end{eqnarray}
where $f_q(\epsilon, \beta)$ [$ \equiv f_q(\epsilon)$] signifies
the generalized distributions, $q$-BED and $q$-FDD, given by
\begin{eqnarray}
f_q(\epsilon, \beta) 
&=& \frac{1}{X_q} \int_0^{\infty}
G\left(u;\frac{q}{q-1}, \frac{1}{(q-1)\beta} \right) \: 
Y_1(u)f_1(\epsilon,u)\:du \nonumber  \\
&& \label{eq:C4}
\hspace{8cm}\mbox{for $q > 1$}, \\
&=& \frac{i}{2 \pi X_q} 
\int_C  H\left(t;\frac{q}{1-q},\frac{1}{(1-q)\beta} \right)
Y_1(-t)\:f_1(\epsilon,- t)\: dt \nonumber \\
&& \label{eq:H4}
\hspace{8cm}\mbox{for $q < 1$},
\end{eqnarray}
with the density of states $\rho(\epsilon)$ given by
\begin{eqnarray}
\rho(\epsilon) &=& \sum_k \delta(\epsilon-\epsilon_k).
\label{eq:C3}
\end{eqnarray}

In order to examine the $(q-1)$ expansion of the generalized
distributions, we set $\hat{O}=\hat{n}_k$ in Eq. (\ref{eq:T22})
where $\hat{n}_k$ denotes the number operator of the state $k$.
A simple calculation leads to the $O(q-1)$ expansion of
the generalized distribution given by
\begin{eqnarray}
f_q(\epsilon,\beta) &=& f_1(\epsilon,\beta)
+ (q-1)\left[\beta \frac{\partial f_1}{\partial \beta}
+ \frac{1}{2}\beta^2 \frac{\partial^2 f_1}{\partial \beta^2} \right]
+ \cdot\cdot, \\
&=& f_1(\epsilon,\beta)
+ (q-1)\left [(\epsilon-\mu) \frac{\partial f_1}{\partial \epsilon}
+ \frac{1}{2}(\epsilon-\mu)^2 \frac{\partial^2 f_1}{\partial \epsilon^2} \right]
+ \cdot\cdot.
\label{eq:T5}
\end{eqnarray}
In deriving Eq. (\ref{eq:T5}), we have employed the relation: 
$(\partial Y_1/\partial \beta)/Y_1(\beta)
=-\langle H-\mu N\rangle_1+(N_q- \mu N_q)
\simeq O(q-1)$.
In Appendix A, we have made a similar analysis with the un-normalized MEM,
showing that Eq. (\ref{eq:T5}) is consistent with Eq. (\ref{eq:V10}) 
which agrees with the result in the AA \cite{Tsallis95}.

\subsubsection{Properties of the generalized distribution}
We will examine some limiting cases of the generalized distribution 
given by Eqs. (\ref{eq:C4}) and (\ref{eq:H4}).

\noindent
\vspace{0.5cm}
(1) In the limit of $q \rightarrow 1.0$, Eq. (\ref{eq:T5}) leads to
\begin{eqnarray}
f_q(\epsilon, \beta) &=& 
f_1(\epsilon, \beta).
\label{eq:C6}
\end{eqnarray}

\noindent
\vspace{0.5cm}
(2) In the zero-temperature limit of $\beta \rightarrow \infty$, 
the $q$-FDD becomes
\begin{eqnarray}
f_q(\epsilon, T=0) 
&=& \Theta(\mu-\epsilon) = f_1(\epsilon, T=0), 
\label{eq:C7}  
\end{eqnarray}
where $\Theta(x)$ stands for the Heaviside function.
Equation (\ref{eq:C7}) implies that the ground-state 
FD distribution is not modified by
the nonextensivity. 

\noindent
\vspace{0.5cm}
(3) In the high-temperature limit of $\beta \rightarrow 0.0$,
where  
$\Omega_1 \simeq - (1/\beta) \sum_k e^{-\beta (\epsilon_k-\mu)}$
with $\ln (1\pm x) \simeq \mp x$ for small $x$, 
we obtain ($\mu=0.0$)
\begin{eqnarray}
f_q(\epsilon, \beta \rightarrow 0) &\propto & 
[1+(q-1)\beta (\epsilon-E_q)]^{\frac{1}{1-q}-1}
= [e_q^{-\beta \:(\epsilon-\mu)}]^q, 
\label{eq:C8}
\end{eqnarray}
$e_q^x$ expressing the $q$-exponential function defined by
\begin{eqnarray}
e_q^{x} &=& \exp_q(x) = [1+(1-q)x]^{\frac{1}{1-q}}
\hspace{1cm}\mbox{for $1+(1-q)x > 0$}, \\
&=& 0
\hspace{4.5cm}\mbox{for $1+(1-q)x \leq 0$},
\label{eq:A7} 
\end{eqnarray}
with the cut-off properties.
Equation (\ref{eq:C8}) corresponds to the escort distribution,
\begin{eqnarray}
P_q(\epsilon) & =& \frac{p_q(\epsilon)^q}{c_q}
\propto  [e_q^{-\beta \:(\epsilon-\mu)}]^q,
\label{eq:C9}
\end{eqnarray}
with the $q$-exponential distribution $p_q(\epsilon)$ given by
\begin{eqnarray}
p_q(\epsilon) &=& e_q^{-\beta \:(\epsilon-\mu)}. 
\label{eq:C10}
\end{eqnarray}

Equations (\ref{eq:C4}) and (\ref{eq:H4}) shows that 
the $\epsilon$ dependence of $f_q(\epsilon, \beta)$ 
arises from that of $f_1(\epsilon, \beta)$.
In particular, the $q$-FDD preserves the same $\epsilon$ symmetry 
as $f_1(\epsilon,\beta)$:

\noindent
(a) $f_q(\epsilon,\beta)=1/2$ for $\epsilon=\mu$,

\noindent
(b) $f_q(\epsilon,\beta)$ has the {\it anti-symmetry}:
\begin{eqnarray}
f_q(-\delta\epsilon+\mu,\beta)-\frac{1}{2}
=\frac{1}{2}-f_q(\delta \epsilon+\mu,\beta)
\hspace{1cm} \mbox{for $\delta \epsilon > 0$},
\label{eq:P1} \nonumber
\end{eqnarray} 

\noindent
(c) $\partial f_q(\epsilon,\beta)/\partial \epsilon$ 
is symmetric with respect to $\epsilon=\mu$.

\subsection{Numerical calculations}

\subsubsection{Model for electrons}

For model calculations of electron systems,
we employ a uniform density of state given by
\begin{eqnarray}
\rho(\epsilon)=(1/2 W) \;\Theta(W- \vert \epsilon \vert),
\label{eq:F1}
\end{eqnarray}
where $W$ denotes a half of the total band width. 
We have performed numerical calculations of $E_q$ and $\mu$ for $q \geq 1.0$
as a function of $T$ for a given number of particles of $N$ 
and the density of states $\rho(\epsilon)$.
We may obtain analytical expressions for 
$\Xi_1(u)$, $N_1(u)$ and $E_1(u)$
which are necessary for our numerical calculations. 
By using Eq. (\ref{eq:F1}) for Eqs. (\ref{eq:B5})-(\ref{eq:B9}), 
we obtain (with $W=1.0$)
\begin{eqnarray}
\Xi_1(u) &=& e^{- u \:\Omega_1(u)}, \nonumber \\
\Omega_1(u) &=& -\frac{1}{2 u}
\{ \ln[1+e^{-u(1-\mu)}] - \ln[1+e^{-u(1+\mu)}] 
+\ln[1+e^{u(1+\mu)}] - \ln[1+e^{u(1-\mu)}] \} \nonumber \\
&-& \frac{1}{2 u^2}\{Li_2(-e^{-u(1+\mu)})
-Li_2(-e^{u(1-\mu)}) \}, \nonumber  \\
N_1(u) &=& 1 +\frac{1}{2 u}[\ln(1+e^{-u(1+\mu)})
-\ln(1+e^{u(1-\mu)})], \nonumber  \\
E_1(u) &=& -\frac{1}{2 u}[\ln(1+e^{-u(1+\mu)})
+\ln(1+e^{u(1-\mu)})] \nonumber \\
&+&\frac{1}{2 u^2}[Li_2(-e^{-u(1+\mu)})
-Li_2(-e^{u(1-\mu)}) ], \nonumber 
\end{eqnarray}
where $Li_n(z)$ denotes the $n$th polylogarithmic function defined by
\begin{eqnarray}
Li_n(z) &=& \sum_{k=1}^{\infty} \frac{z^k}{k^n}.\nonumber 
\end{eqnarray}

We adopt $N=0.5$, for which $\mu=0.0$ independent of 
the temperature because of the adopted uniform density of states given 
by Eq. (\ref{eq:F1}).
The temperature dependence of $E_q$
calculated self-consistently from Eqs.(\ref{eq:B14})-(\ref{eq:B16}), 
is shown in Fig. \ref{fig1} whose inset shows the enlarged plot
for low temperatures ($k_B T/W \lesssim 0.1$). 
We note that $E_q$ at low temperatures is larger for
larger $q$ although this trend is reversed at higher
temperatures ($k_B T \gtrsim 0.3 $).

The calculated $q$-FDDs $f_q(\epsilon)$
for various $q$ values for $k_B T/W=0.1$ are shown in
Figs. \ref{fig2} (a) and \ref{fig2} (b) whose ordinates are 
in the linear and logarithmic scales, respectively.
It is shown that with more increasing $q$ from unity, $f_q(\epsilon)$
at $\epsilon \gg \mu$ has a longer tail.
The properties of $f_q(\epsilon)$ are more clearly seen in
its derivative of $-\partial f_q(\epsilon)/\partial \epsilon$,
which is plotted in Fig. \ref{fig3} with the logarithmic ordinate.
We note that $-\partial f_q(\epsilon)/\partial \epsilon$ is symmetric
with respect of $\epsilon=\mu$.
With increasing $q$ above unity, 
$-\partial f_q(\epsilon)/\partial \epsilon$ has a longer tail.
Dotted and solid curves for $q < 1.0$ in Figs. \ref{fig2} and \ref{fig3}
will be discussed in Sec. 3.3.

\subsubsection{The Debye model for phonons}

We adopt the Debye model whose phonon density of states
is given by
\begin{eqnarray}
\rho(\omega) &=& A \: \omega^2 
\hspace{1cm} \mbox{for $0 < \omega \leq \omega_D$},
\label{eq:G1}
\end{eqnarray}
where $A= 9 N_a/w_D^3$, $N_a$ denotes the number
of atoms, $\omega$ the phonon frequency
and $\omega_D$ the Debye cutoff frequency.
By using Eq. (\ref{eq:G1}) to Eqs. (\ref{eq:B5})-(\ref{eq:B9}), 
we may obtain (with $\omega_D=1.0$ and $\mu=0$),
\begin{eqnarray}
\Xi_1(u) &=& e^{- u \: \Omega_1(u)}, \nonumber  \\
\Omega_1(u) &=& \frac{A}{180 u} \left[\frac{4 \pi^4}{u^3}+15 u
-60 \ln(1-e^{u})
+60 \ln(1-\cosh u+\sinh u ) \right] \nonumber \\
&-& \frac{A}{u^4}[u^2 \:Li_2(e^{u})
- 2 u\: Li_3(e^{u})+ 2 Li_4(e^{u})], \nonumber \\
N_1(u) &=& - \frac{A}{3 u^3}
[ u^3 - 3 u^2 \ln(1-e^{u})
- 6 u Li_2(e^{u}) + 6 Li_3(e^{u}) - 6 \:\zeta(3) ], \nonumber \\
E_1(u) &=& A \left[\frac{\ln(1-e^{u})}{u}
+ \frac{3 Li_2(e^{u})}{u^2}
- \frac{6 Li_3(e^{u})}{u^3}
+ \frac{6 Li_4(e^{u})}{u^4}
- \frac{1}{4} - \frac{\pi^4}{15 u^4} \right]. \nonumber 
\end{eqnarray}

We have performed numerical calculations with the Debye model for $q \geq 1.0$.
The temperature dependence of self-consistently calculated $E_q$ 
is shown in Fig. \ref{fig4} where
inset shows the enlarged plots for low temperatures ($T/T_D < 0.5$).
We note that $E_q$ at low temperatures is larger for larger $q$.

The calculated $q$-BEDs $f_q(\epsilon)$
for various $q$ values for $T/T_D=0.01$ are shown in
Fig. \ref{fig5} whose ordinate is in the logarithmic scale:
they are indistinguishable in the linear scale.
It is shown that with more increasing $q$, $f_q(\epsilon)$
at $\epsilon \gg \mu$ has a longer tail.
Dotted and solid curves for $q < 1.0$ will be discussed
in Sec. 3.3.

\section{THE INTERPOLATION APPROXIMATION}

\subsection{Analytic expressions of the generalized distributions}

In the preceding Sec. 2,
we have discussed the generalized distributions based on 
the exact representation given by Eqs. (\ref{eq:C4}) and (\ref{eq:H4}).
It is, however, difficult to calculate them because
they need self-consistent calculations of $N_q$ and $E_q$.
If we assume 
\begin{eqnarray}
\left( \frac{1}{X_q} \right)
e^{u(E_q-\mu N_q)}  \Xi_1(u) &=& 1,
\label{eq:C7b}
\end{eqnarray}
in Eqs. (\ref{eq:C4}) and (\ref{eq:H4}), we obtain
the approximate generalized distributions 
given by
\begin{eqnarray}
f_q^{IA}(\epsilon, \beta) &=&
\int_0^{\infty} G\left(u;\frac{q}{q-1}, \frac{1}{(q-1)\beta} \right)
\:f_1(\epsilon,u)\:du.
\hspace{0.5cm} \mbox{for $q > 1.0$},
\label{eq:M1} \\
&=& 
\frac{i}{2 \pi} 
\int_C  H\left(t;\frac{q}{1-q},\frac{1}{(1-q)\beta} \right)
\:f_1(\epsilon,-t)\: dt 
\hspace{0.5cm} \mbox{for $q < 1.0$},
\label{eq:M2} 
\end{eqnarray}
where $G(u;a,b)$ and $H(t;a,b)$ are given by 
Eqs. (\ref{eq:B7}) and (\ref{eq:H6}), respectively.
Equations (\ref{eq:M1}) and (\ref{eq:M2}) are referred to 
as the {\it interpolation approximation} (IA) in this paper because 
they have the important interpolating character,
as will be shown shortly (Sec. 3.2). 
Note that calculations of $f_q^{IA}(\epsilon, \beta)$ by
Eqs. (\ref{eq:M1}) and (\ref{eq:M2}) do not require $N_q$ and $E_q$.
Equation (\ref{eq:M1}) may be regarded as a kind of the SA. 

One of advantages of the IA is that we can obtain the simple 
analytic expressions for the $q$-BED and $q$-FDD as follows.

\noindent
{\bf (1) $q$-BED}

We first
expand the Bose-Einstein distribution $f_1(\epsilon, \beta)$ as
\begin{eqnarray}
f_1(\epsilon,\beta)
&=&\sum_{n=0}^{\infty} \: e^{-(n+1) x}
\hspace{1cm}\mbox{for $x > 0 $}, 
\label{eq:M10}
\end{eqnarray}
where $x=\beta(\epsilon-\mu)$.
Substituting Eq. (\ref{eq:M10}) to Eqs. (\ref{eq:M1}) and (\ref{eq:M2}),
and employing Eq. (\ref{eq:B1}) and (\ref{eq:H1}),  
we obtain the $q$-BED in the IA given by
\begin{eqnarray}
f_q^{IA}(\epsilon,\beta) 
&=& \sum_{n=0}^{\infty}\:[e_q^{-(n+1)\:x}]^q
\hspace{3cm}\mbox{for $0 < q < 3 $}, 
\label{eq:M12}\\
&=& \left[ \frac{1}{(q-1) x} \right]^{\frac{q}{q-1}}
\zeta\left(\frac{q}{q-1}, \frac{1}{(q-1) x} +1 \right) 
\hspace{0.5cm}\mbox{for $1 < q < 3$},
\label{eq:N7} 
\end{eqnarray}
where $\zeta(z,a)$ denotes the Hurwitz zeta function:
\begin{eqnarray}
\zeta(z,a)&=&\sum_{k=0}^{\infty}\:\frac{1}{(k+a)^{z}}
= \frac{1}{\Gamma(z)}\int_{0}^{\infty} \: \frac{t^{z-1} e^{-a t}}{1-e^{-t}}\:dt
\hspace{0.5cm}\mbox{for $ \Re \:z > 1$}. \nonumber \label{eq:N5} 
\end{eqnarray}
It derivative is given by
\begin{eqnarray}
\frac{\partial f_q^{IA}}{\partial x} 
&=& -\sum_{n=0}^{\infty} \:q(n+1)[e_q^{-(n+1) x }]^{(2q-1)}
\hspace{1cm}\mbox{for $0 < q < 3 $}.
\end{eqnarray}
We may easily realize that
$f_q(\epsilon, \beta)$ in Eq. (\ref{eq:M12})
reduces to $f_1(\epsilon, \beta)$
in the limit of $q \rightarrow 1.0$ where $e_q^x \rightarrow e^x$.

\vspace{0.5cm}
\noindent
{\bf (2) $q$-FDD}

The Fermi-Dirac distribution $f_1(\epsilon,\beta)$ 
may be expanded as
\begin{eqnarray}
f_1(\epsilon,\beta) 
&=& \sum_{n=0}^{\infty} \: (-1)^n \:e^{-(n+1) x}
\hspace{1cm}\mbox{for $x > 0$}, 
\label{eq:M4} \\
&=& \frac{1}{2}
\hspace{4cm}\mbox{for $x = 0$}, \\
&=& \sum_{n=0}^{\infty} \: (-1)^n \:e^{-n \:\vert x \vert}
\hspace{1cm}\mbox{for $x < 0$}, 
\label{eq:M5}
\end{eqnarray}
where $x=\beta(\epsilon-\mu)$.
Substituting Eqs. (\ref{eq:M4})-(\ref{eq:M5}) to Eqs. (\ref{eq:M1}) 
and (\ref{eq:M2}), and employing Eq. (\ref{eq:B1}) and (\ref{eq:H1}), 
we obtain the $q$-FDD in the IA given by
\begin{eqnarray}
f_q^{IA}(\epsilon, \beta) 
&=& F_q(x)
\hspace{4cm} \mbox{for $ x  > 0$},
\label{eq:N1} \\
&=& \frac{1}{2}
\hspace{5cm} \mbox{for $ x = 0$}, \\
&=& 1 - F_q(\vert x \vert )
\hspace{3cm} \mbox{for $ x < 0 $},
\label{eq:N2}
\end{eqnarray}
with
\begin{eqnarray}
F_q(x) &=& 
\sum_{n=0}^{\infty} \: (-1)^n \:[e_q^{-(n+1) x}]^q
\hspace{2cm}\mbox{for $0 < q < 3$}, 
\label{eq:M7} \\
&=&\left[ \frac{1}{2(q-1) x} \right]^{\frac{q}{q-1}}
\{ \zeta\left(\frac{q}{q-1}, \frac{1}{2(q-1) x}+\frac{1}{2} \right)
\nonumber \\
&& - \zeta\left(\frac{q}{q-1}, \frac{1}{2(q-1) x}+1 \right) 
\}
\hspace{1cm}\mbox{for $1 < q < 3$}.\label{eq:N3} 
\end{eqnarray}
It derivative is given by
\begin{eqnarray}
\frac{\partial f_q^{IA}}{\partial x} 
&=& -\sum_{n=0}^{\infty} (-1)^n \:q(n+1)[e_q^{-(n+1)\: \vert x \vert }]^{(2q-1)}
\hspace{1cm}\mbox{for $0 < q < 3 $},
\end{eqnarray}
which is symmetric with respect to $x=0$.
The $q$-FDD given by Eqs. (\ref{eq:N1})-(\ref{eq:M7}) 
reduces to $f_1(\epsilon, \beta)$ in the limit of $q \rightarrow 1.0$. 

We may obtain a useful expression of the $q$-FDD
for $\vert x \vert < 1$ given by (see Appendix B.1)
\begin{eqnarray}
f_q^{IA} &\simeq& \frac{1}{2}-\frac{q}{4}\: x 
+ \frac{q (2q-1)(3q-2)}{48} x^3 +\cdot\cdot, 
\label{eq:R1} \\
\frac{\partial f_q^{IA}}{\partial x}
&\simeq& -\frac{q}{4} + \frac{q (2q-1)(3q-2)}{16} x^2 +\cdot\cdot
\hspace{1cm}\mbox{for $ 0 < q < 3 $}.
\label{eq:R2}
\end{eqnarray}

In the case of $q < 1.0$, summations over $n$ 
in the $q$-BED and $q$-FDD [Eqs. (\ref{eq:M12}) and (\ref{eq:M7})] are
terminated when the condition: $n+1 > 1/(1-q)x$ is satisfied
because of the cut-off properties of the $q$-exponential function
given by Eq. (\ref{eq:A7}). 
Then the $q$-FDD for $q < 1.0$ has the cut-off properties given by
\begin{eqnarray}
f_q^{IA}(\epsilon) &=& 0.0
\hspace{1cm}\mbox{for $\epsilon-\mu > 1/(1-q)\beta$}, 
\label{eq:S1} \\
&=& 1.0
\hspace{1cm}\mbox{for $\epsilon-\mu < -1/(1-q)\beta$},
\label{eq:S2}
\end{eqnarray} 
while the $q$-BED has the cut-off properties
given by Eq. (\ref{eq:S1}). These are the same as the $q$-exponential 
distribution $p_q(\epsilon)$ given by Eq. (\ref{eq:C10}).

\subsection{Comparison with the exact approach}

From Eqs. (\ref{eq:T20}) and (\ref{eq:T21}) with $Y_1(u)=1.0$,
the $q$-BED and $q$-FDD for $q \simeq 1.0$ in the IA become
\begin{eqnarray}
f_q^{IA}(\epsilon,\beta) &=& f_1(\epsilon,\beta)
+ (q-1) \left[ (\epsilon-\mu)  \frac{\partial f_1}{\partial \epsilon}
+ \frac{1}{2} (\epsilon-\mu)^2 
\frac{\partial^{2} f_1}{\partial \epsilon^2}\right]
+\cdot\cdot,
\label{eq:K12} 
\end{eqnarray}
which is in agreement with those in 
the EA given by Eq. (\ref{eq:T5}) within $O(q-1)$. 
In the zero-temperature limit, the $q$-FDD reduces to
\begin{eqnarray}
f_q^{IA}(\epsilon,T=0) = \Theta(\mu-\epsilon).
\label{eq:Q1}
\end{eqnarray}
In the opposite high-temperature limit, the $q$-BED and $q$-FDD
become
\begin{eqnarray}
f_q^{IA}(\epsilon,\beta \rightarrow 0) \propto [e_q^{-x}]^{q}.
\label{eq:Q2}
\end{eqnarray}
Equations (\ref{eq:Q1}) and (\ref{eq:Q2}) agree
with Eqs. (\ref{eq:C7}) and (\ref{eq:C8}), respectively, for the EA.
Thus the generalized distributions in the IA have the interpolation
properties, yielding results
in agreement with those in the EA within $O(q-1)$ and
in high- and low-temperature limits.

\subsection{Numerical calculations}
Numerical calculations of $f_q^{IA}(\epsilon,\beta)$
[$\equiv f_q^{IA}(\epsilon)$] have been performed. 
Results of the FDD of $f_q^{EA}(\epsilon)$ in the EA 
for $q > 1.0$ and $k_B T/W=1.0$ 
are shown in Fig. \ref{fig6}.
With more increasing $q$, the distributions have longer tails,
as shown in Fig. \ref{fig2} for $k_B T/W=0.1$.
The result in the IA is in good agreement with the EA 
because the ratio defined by
$\lambda \equiv f_q^{IA}(\epsilon)/f_q^{EA}(\epsilon)$
is $0.97 \lesssim \lambda \lesssim 1.01$
for $-10 < \epsilon <10$ as shown in the inset. 
The $\epsilon$ dependence of the BED of $f_q^{EA}(\epsilon)$ in the EA 
for $q > 1.0$ and $T/T_D=0.1$ is plotted in 
Fig. \ref{fig7} which shows similar behavior to those
for $T/T_D=0.01$ shown in Fig. \ref{fig6}.
Its inset shows that the ratio of $\lambda$
is $0.7 \lesssim \lambda \lesssim 1.0$ for $1.0 < q \leq 1.2$. 
These calculations justify, to some extent, the distribution in the IA given 
by Eqs. (\ref{eq:N7}), (\ref{eq:N1})-(\ref{eq:N2}) and (\ref{eq:N3}).

We have calculated the $q$-BED and $q$-FDD also for $q < 1.0$, by using
Eqs. (\ref{eq:M12}), (\ref{eq:N1})-(\ref{eq:M7}).
Dotted and solid curves in Fig. \ref{fig2} 
show the $q$-FDD of $f_q^{IA}(\epsilon)$ for $q=0.9$ and $q=0.8$, respectively.
Their derivatives of $-\partial f_q^{IA}(\epsilon)/\partial \epsilon $
for $q=0.9$ and $q=0.8$ are plotted by the dotted and solid 
curves, respectively, in Fig. \ref{fig3}.
Dotted and solid curves in Fig. \ref{fig5} show the $q$-BED of
$f_q^{IA}(\epsilon)$ for $q=0.9$ and $q=0.8$, respectively.
With more decreasing $q$ from unity,
the curvature of $f_q(\epsilon)$ in both 
$q$-BED and $q$-FDD become more significant.
The cut-off properties in the $q$-FDD and $q$-BED 
given by Eqs. (\ref{eq:S1}) and (\ref{eq:S2})
are realized in Figs. \ref{fig2} and \ref{fig5}.
We expect that $f_q^{IA}(\epsilon)$ in the case of $q < 1.0$
is a good approximation of the $q$-BED and $q$-FDD 
as in the case of $q > 1.0$.

\section{DISCUSSION}
\subsection{Comparison with previous studies}
It is interesting to compare our results to
those previously obtained with some approximations. 

\vspace{0.5cm}
\noindent
(A) {\it The factorization approximation}

B\"{u}y\"{u}kkilic, Demirhan and G\"{u}lec \cite{Buy95} 
derived the $q$-BED and $q$-FDD given by
\begin{eqnarray}
f_q^{FA}(\epsilon,\beta) &=& 
\frac{1}{\{e_q[-\beta (\epsilon-\mu)] \}^{-1}\mp 1},
\label{eq:K1}
\end{eqnarray} 
adopting the FA given by 
\begin{eqnarray}
Q &=& [1-(1-q)\sum_{n=1}^N x_n]^{\frac{1}{1-q}}, 
\label{eq:K0} \\
&\simeq& \prod_{n=1}^N [1-(1-q) x_n]^{\frac{1}{1-q}},
\label{eq:K2}
\end{eqnarray}
to evaluate the grand canonical partition function,
the upper (lower) sign in Eq. (\ref{eq:K1}) being applied to boson (fermion).

It is noted that if we assume the factorization approximation:
$[e_q^{-(n+1)x}]^q \simeq (e_q^{-x})^q [(e_q^{-x})^q]^n$
in $f_q^{IA}(\epsilon)$ [Eqs. (\ref{eq:M12}) and (\ref{eq:M7})], 
we obtain
\begin{eqnarray}
f_q(\epsilon,\beta) &\simeq& 
\frac{1}{\{e_q[-\beta (\epsilon-\mu)] \}^{-q}\mp 1},
\end{eqnarray}
which is similar to Eq. (\ref{eq:K1}) \cite{Nunes01,Martinez03}.

\vspace{0.5cm}
\noindent
(B) {\it The superstatistical approximation}

In the SA, the generalized distribution is expressed as a superposition
of $f_1(\epsilon)$ \cite{Wilk00,Beck01},
\begin{eqnarray}
f_q^{SA}(\epsilon, \beta) &=&
\int_0^{\infty} G\left(u;\frac{1}{q-1}, \frac{1}{(q-1)\beta} \right)
\:f_1(\epsilon,u)\:du,
\label{eq:K3}
\end{eqnarray}
which is similar to but different from $f_q^{IA}(\epsilon,\beta)$
given by Eq. (\ref{eq:M1}).
Recently the $q$-FDD equivalent to Eq. (\ref{eq:K1}) is obtained
by employing the SA in a different way \cite{Hasegawa09}.

The properties of the
generalized distributions of the EA, IA, FA and SA
in the limits of $q \rightarrow 1.0$, $\beta \rightarrow \infty$
and $\beta \rightarrow 0.0$ are compared in Table 1.
The result of the IA agrees with that of the EA within $O(q-1)$
as mentioned before.
However, the $O(q-1)$ contributions in the FA and SA are 
different from that in the EA.
In the zero-temperature limit, all the $q$-FDDs reduce to
$\Theta(\mu-\epsilon)$. In the opposite high-temperature limit,
the generalized distributions in the FA and SA
reduce to $e_q^{-\beta \epsilon}$, while those
in the EA and IA become $[e_q^{-\beta \epsilon}]^q$
where the power index $q$ arises from the escort probability
in the OLM-MEM given by Eq. (\ref{eq:C9}) \cite{Martinez00,Ferri05}.

Figure \ref{fig8} shows $q$-BED for $q=1.1$ and $q=1.2$ calculated 
by the FA, SA and EA with the logarithmic ordinate. 
For a comparison, we show $f_q(\epsilon)$ for $q=1.0$ by dashed curves.
The difference among $f_q(\epsilon)$'s of the three methods
is clearly realized:
tails in the $q$-BED of the FA and SA are overestimated.

Figure \ref{fig9} shows $q$-FDD for $q=1.1$ and $q=1.2$
calculated by the EA, FA and SA
with the logarithmic ordinate (for more detailed $f_q^{FA}(\epsilon)$,  
see Fig. 1 of Ref. \cite{Hasegawa09}).
Tails in the FA and SA have larger than that in the EA,
as in the case of the $q$-BED shown in Fig. \ref{fig8}. 

Figures \ref{fig10}(a) and 10(b) show the $q$-FDD and its derivative,
respectively, calculated in the IA and FA.
For $q=0.9$, $f_q^{FA}(\epsilon)$ at $ \epsilon < \mu$
is much reduced than $f_q^{IA}(\epsilon)$.
For $q=1.1$, on the contrary, $f_q^{FA}(\epsilon)$ at $ \epsilon > \mu$
is much increased than $f_q^{IA}(\epsilon)$.
These lead to an overestimate of electron excitations across 
the fermi level $\mu$ in the FA. Furthermore
$-\partial f_q^{FA}(\epsilon)/\partial \epsilon$ in the FA is not
symmetric with respect to $\epsilon=\mu$ in contrast to that in the IA.

The FA was criticized in Refs. \cite{Pennini95}\cite{Lenzi01}
but justified in Ref. \cite{Wang97}.
The dismissive study \cite{Pennini95}
was based on a simulation with $N=2$. 
In contrast, the affirmative study \cite{Wang97}
performed simulations with $N=10^5$ and $10^{15}$.
Lenzi, Mendes, da Silva and Malacarne \cite{Lenzi01} criticized the FA,
applying  the EA \cite{Rajagopal98,Lenzi99}
to independent harmonic oscillators with $N \leq 100$. 
Our results are consistent with Refs. \cite{Pennini95,Lenzi01}. 
The FA given by Eq. (\ref{eq:K2})
has been explicitly or implicitly 
employed in many studies not only for quantum 
but also classical nonextensive systems.
It would be necessary to examine the validity of
these studies using the FA from the viewpoint of the 
exact representation \cite{Rajagopal98,Lenzi99,Note1}. 

By using Eqs.  (\ref{eq:B1}) and (\ref{eq:H1}), 
we may rewrite $Q$ in Eq. (\ref{eq:K0}) as
\begin{eqnarray}
Q &=& [1-(1-q)x_1]^{\frac{1}{1-q}} 
\otimes_q \cdot\cdot
\otimes_q [1-(1-q)x_N]^{\frac{1}{1-q}},
\label{eq:K4} \\
&=& \int_0^{\infty} \: G\left( u;\frac{1}{q-1},\frac{1}{q-1} \right)
\prod_{n=1}^N e^{- u\: x_n}\:du
\hspace{1cm}\mbox{for $q > 1.0$},
\label{eq:K5} \\
&=& \frac{i}{2 \pi}\int_C H\left(t;\frac{1}{1-q} \right)
\prod_{n=1}^N e^{(1-q) \:t \:x_n}\:dt
\hspace{2cm}\mbox{for $q < 1.0$},
\label{eq:K6}
\end{eqnarray}
where $\otimes_q$ denotes the $q$-product defined by
\cite{Borges04}
\begin{eqnarray}
x \otimes_q y &\equiv& [x^{1-q}+y^{1-q}-1]^{\frac{1}{1-q}}.
\end{eqnarray}
Equations (\ref{eq:K5}) and (\ref{eq:K6}) are the integral 
representations of the $q$-product given by Eq. (\ref{eq:K4}).
The result of the FA in (\ref{eq:K2}) is derived if we may 
exchange the order of integral and product in Eqs. (\ref{eq:K5}) 
and (\ref{eq:K6}), which is of course forbidden. 

\subsection{The generalized Sommerfeld expansion}

We will investigate the generalized Sommerfeld expansion 
for an arbitrary function
$\phi(\epsilon)$ with the $q$-FDD of $f_q(\epsilon)$
given by \cite{Hasegawa09}
\begin{eqnarray}
I &=& \int \phi(\epsilon) f_q(\epsilon) \:d\epsilon, \\
&=& \int^{\mu} \phi(\epsilon) \: d \epsilon
+ \sum_{n=1}^{\infty} c_{n,q} \:(k_B T)^n \:\phi^{(n-1)}(\mu),
\label{eq:E6} 
\end{eqnarray}
with
\begin{eqnarray}
c_{n,q}&=& - \:\frac{\beta^n}{n!} 
\int (\epsilon-\mu)^n \frac{\partial f_q(\epsilon)}
{\partial \epsilon} \:d \epsilon.
\label{eq:E2}
\end{eqnarray}
Substituting $f_q(\epsilon)$ in the EA given by 
Eq. (\ref{eq:T5}) to Eq. (\ref{eq:E2}), 
and using integrals by part, we obtain $c_{n,q}$ for even $n$,
\begin{eqnarray}
\frac{c_{n,q}^{EA}}{c_{n,1}} &= & 
1+ \frac{n(n-1)}{2}(q-1)+\cdot\cdot 
\hspace{1cm}\mbox{for even $n$}, 
\label{eq:T6}\\
&=& 1+(q-1)+\cdot\cdot \hspace{1cm} \mbox{for $n=2$}, 
\label{eq:T13} \\
&=& 1+6(q-1)+\cdot\cdot \hspace{1cm} \mbox{for $n=4$},
\label{eq:T14}
\end{eqnarray}
while $c_{n,q}=0$ for odd $n$,
where $c_{n,1}$ denotes the relevant expansion coefficient for $q=1.0$:
$c_{2,1}=\pi^2/6$ (=1.645) and $c_{4,1}=7 \pi^4/360$ (=1.894)
{\it et. al.}. 
Equation (\ref{eq:T6}) shows that $c_{n,q}$ is increased with increasing $q$.

By using $f_q^{IA}(\epsilon)$ in the IA,
we may obtain $c_{n,q}$ given by (for details, see Appendix B.2)
\begin{eqnarray}
\frac{c^{IA}_{n,q}}{c_{n,1}} 
&=& \frac{\Gamma(\frac{1}{q-1}+1-n)}
{(q-1)^n \Gamma(\frac{1}{q-1}+1)}
\hspace{1cm}\mbox{for even $n$, $q>1$ }, \\
&=& \frac{\Gamma(\frac{q}{1-q}+1)}
{(1-q)^n \:\Gamma(\frac{q}{1-q}+1+n)} 
\hspace{1cm}\mbox{for even $n$, $q <1$}, \\
&=& \frac{1}{2-q}
\hspace{4cm}\mbox{for $n=2$}, 
\label{eq:T11} \\
&=& \frac{1}{(2-q)(3-2q)(4-3q)}
\hspace{1cm}\mbox{for $n=4$}.
\label{eq:T12}  
\end{eqnarray}
It is easy to see that Eqs. (\ref{eq:T11}) and (\ref{eq:T12}) 
are in agreement with Eq. (\ref{eq:T13}) and (\ref{eq:T14}),
respectively, of the EA within $O(q-1)$.

A simple calculation using $f_q^{SA}(\epsilon)$ leads to
\begin{eqnarray}
\frac{c_{n,q}^{SA}}{c_{n,1}}
&=& \frac{\Gamma(\frac{1}{q-1}-n)}
{(q-1)^n\:\Gamma(\frac{1}{q-1})} 
\hspace{1cm}\mbox{for even $n$ ($q > 1$)},\\
&=& \frac{1}{(2-q)(3-2q)}
\hspace{1cm}\mbox{for $n=2$}, \nonumber \\
&=& \frac{1}{(2-q)(3-2q)(4-3q)(5-4q)}
\hspace{1cm}\mbox{for $n=4$}, \nonumber 
\end{eqnarray}
which are similar to those given by Eqs. (\ref{eq:T11}) and (\ref{eq:T12}).

The Sommerfeld expansion coefficients
in the FA may be calculated with the use of $f_q^{FA}(\epsilon)$ \cite{Hasegawa09}.
A comparison among the $O(q-1)$ contributions to $c_{n,q}$ ($n=1-4$)
in the four methods of EA, IA, FA and SA is made in Table 2.
The results of the IA coincide with those of the EA.
The $O(q-1)$ contributions to $c_{2,q}$ and $c_{4,q}$ in the SA 
are three and $5/3$ times larger, respectively, than those in the EA.
The $O(q-1)$ contributions to $c_{2,q}$ and $c_{4,q}$
in the FA are vanishing. 
It is noted that $c_{1,q}^{FA} \neq 0$ and $c_{3,q}^{FA} \neq 0$
in contrast with the results of
$c_{1,q}= c_{3,q}= 0$ in the EA, IA and SA. 
This is due to a lack of the symmetry in 
$-\partial f_q^{FA}(\epsilon)/\partial \epsilon$ 
with respect to $\epsilon=\mu$ as shown 
in Fig. \ref{fig10}(b).

Figure \ref{fig11}(a) shows the $q$ dependence of
coefficients of $c_{n,q}/c_{n,1}$ for $n=2$ and 4 calculated 
by the four methods.
Circles and squares express $c_{n,q}^{EA}$ for $n=2$ and 4, respectively,
calculated by the EA for $k_B T/W=0.1$ (Fig. \ref{fig1}). Solid curves express
$c_{n,q}^{IA}$ in the IA.  
The coefficient for $n=2$ ($n=4$) in the IA is in good agreement
with the result in the EA for $1.0 \leq q \lesssim 1.5$ ($1.0 \leq q \lesssim 1.2$).
$c_{n,q}^{SA}$ shown by chain curves
are overestimated compared to $c_{n,q}^{EA}$ and $c_{n,q}^{IA}$.
Dashed curves denoting $c_{n,q}^{FA}$ \cite{Hasegawa09}
are plotted only for $0.8 \leq q \leq 1.2$, because the FA is
considered to be valid for a small $\vert q-1 \vert $ \cite{Tirnakli00}.
The $q$ dependence of $c_{n,q}^{FA}$ is qualitatively different from
those of the EA, IA and SA:
$c_{n,q}^{FA}$ is symmetric with respect to 
$q=1.0$ whereas those in other three methods are monotonously increased
with increasing $q$.

The energy of electron systems at low temperatures
may be calculated with the use of the generalized Sommerfeld expansion.
By using Eqs. (\ref{eq:E6}) and (\ref{eq:T6}) for Eq. (\ref{eq:F1}) 
with $\phi(\epsilon)= \epsilon \rho(\epsilon)$,
we obtain the energy given by
\begin{eqnarray}
E_q(T) \simeq E_q(0) + c_{2,q} (k_B T)^2 \rho(\mu) + \cdot\cdot,
\label{eq:F2}
\end{eqnarray}
from which the low-temperature 
electronic specific heat is given by
\begin{eqnarray}
C_q(T) &\simeq& \gamma_q T+ \cdot \cdot,
\label{eq:F3}
\end{eqnarray}
with
\begin{eqnarray}
\frac{\gamma_q}{\gamma_1} &=& \frac{c_{2,q}}{c_{2,1}},
\label{eq:F4} \\
\gamma_1 &=& \frac{\pi^2}{3} k_B^2 \rho(\mu),
\end{eqnarray}
where $\gamma_1$ is the linear-$T$ expansion coefficient for $q=1.0$.

The inset of Fig. \ref{fig1} shows that
the calculated energy $E_q$ at low temperatures in the electron model
is larger for a larger $q$, which is  
consistent with larger $\gamma_q$ and $c_{2,q}$
for a larger $q$ as shown in Fig. \ref{fig11}(a).

\subsection{Low-temperature phonon specific heat}

We consider the phonon specific heat at low temperatures.
By using Eqs. (\ref{eq:C2}) and (\ref{eq:T5}), we obtain 
\begin{eqnarray}
C_q &\simeq & \alpha_q T^3+\cdot\cdot ,
\label{eq:G8}
\end{eqnarray}
with
\begin{eqnarray}
\frac{\alpha_q^{EA}}{\alpha_1} &=& 1+6(q-1) + \cdot\cdot, 
\label{eq:G9} \\
\alpha_1 &=& \left( \frac{12 \pi^4}{5} \right) N_a k_B,
\end{eqnarray}
where $\alpha_1$ is the relevant coefficient for $q=1.0$.

The coefficients of low-temperature phonon specific heat $\alpha_q$ 
in the IA, SA and FA are given by (for details, see Appendix B.3)
\begin{eqnarray}
\frac{\alpha_q^{IA}}{\alpha_1} &=& 
\frac{1}{(2-q)(3-2q)(4-3q)},
\label{eq:G10} \\ 
\frac{\alpha_q^{SA}}{\alpha_1} &=& 
\frac{1}{(2-q)(3-2q)(4-3q)(5-4q)}, \\
\frac{\alpha_q^{FA}}{\alpha_1} &=& 1 + O((q-1)^2),
\end{eqnarray}
where the $O(q-1)$ contribution to $\alpha_q^{FA}$ is vanishing 
\cite{Hasegawa09}.
Equation (\ref{eq:G10}) shows that $\alpha_q^{IA}$ agrees with
$\alpha_q^{EA}$ within $O(q-1)$ and that the $\alpha_q^{IA}$ is related
with $c_{4,q}^{IA}$ as
$\alpha_q^{IA}/\alpha_1=c_{4,q}^{IA}/c_{4,1}$.

Coefficients of $\alpha_q/\alpha_1$ calculated by the four methods
are plotted as a function of $q$ in Fig.\ref{fig11}(b).
Squares denote the result of numerical calculation
by the EA for $T/T_D=0.01$ (Fig. \ref{fig4}).
The solid curve express $\alpha_q^{IA}$ which is in good
agreement with the result of the EA for $1.0 \leq q \lesssim 1.2$
but deviates from it at $q \gtrsim 1.2$.
Dashed and chain curves show 
$\alpha_q$ calculated by the FA and SA, respectively.
It is interesting that the result of the SA nearly coincides with
that of the FA for $1.0 \leq q \lesssim 1.2$, where both the results 
of the SA and FA are overestimated compared to the EA.
The inset of Fig. \ref{fig4} shows that the energy $E_q$ 
at low temperatures in the Debye model is larger for 
larger $q$, which is consistent with the $q$-dependence of $\alpha_q$
shown in Fig. \ref{fig11}(b).

\section{CONCLUDING REMARKS}

It is well known that in nonextensive classical statistics, 
the nonextensivity arises from the long-range interaction, 
long-time memory and a multifractal-like space-time \cite{Tsallis04}.
The metastable state or quasi-stationary state
is characterized by long-range interaction
and/or fluctuations of intensive quantities ({\it e.g.,} the inverse
temperature) \cite{Beck07}.
For example, in the long-range-interacting
gravitating systems, the physical quantities are not extensive:
the velocity distribution obeys the power law and
the stable equilibrium state is lacking, which
lead to negative specific heat \cite{Dauxois02}.
The situation is the same also in nonextensive quantum statistics.
It has been reported that the observed black-body radiation
may be explained by the nonextensivity of the order of 
$\vert q-1 \vert \sim 10^{-4}-10^{-5}$ which is attributed to the
long-range Coulomb interaction \cite{Tsallis95}.
Memory effect and long-range interaction cannot be neglected
in weakly non-ideal plasma of stellar core \cite{Lavagno07}.
In addition to the large systems where the interactions may be truly
long range,
one should consider small systems where the range of the interactions 
is of the order of the system size.
Small-size systems would not be extensive, and many
similarities with the long-range case will be realized.
Indeed, the negative specific heat is observed in 
147 sodium clusters \cite{Schmidt01}.
Magnetic properties in nano-magnets may be different from
those in large-size ones \cite{Hasegawa06}.
Small drops of quantum fluids may undergo a Bose-Einstein condensation.
Thanks to recent development in the evaporation cooling technique,
it becomes possible to study Bose-Einstein condensation
in an extremely diluted fluid where the long-range
interactions play essential roles in the condensate stability.
Artificial sonic or optical black hole \cite{Unruh81,Gordon23}
represents an intrigue quantum catastrophic phenomenon.
Only little is known about the thermodynamics of these quantum systems.
Experimental and theoretical studies on these subjects
deepen our understanding of basic quantum phenomena.

To summarize, we have discussed the generalized distributions
of $q$-BED and $q$-FDD in nonextensive quantum statistics based on the EA  
\cite{Rajagopal98,Lenzi99} and IA.
Results obtained are summarized as follows: 

\noindent
(i) with increasing $q$ above $q=1.0$, the $q$-BED and $q$-FDD
have long tails, while they have compact
distributions with decreasing $q$ from unity, 

\noindent
(ii) the coefficients in the generalized Sommerfeld expansion,  
the linear-$T$ coefficient of electronic specific heat and 
the $T^3$ coefficient of phonon specific heat are increased
with increasing $q$ above unity,
whereas they are decreased with decreasing $q$ below unity,

\noindent
(iii) the $O(q-1)$ contributions in the EA agree with those 
in the AA based on the OLM-MEM \cite{Martinez00}
as well as the un-normalized MEM \cite{Curado91}, and 

\noindent
(iv) the generalized distributions given
by simple expressions in the IA proposed in this study
yield results in agreement with those obtained by the EA
within $O(q-1)$ and high- and low-temperature limits. 

\noindent
As for the item (iv), the $q$-BED and $q$-FDD in the IA 
are expected to be useful 
and to play important roles in the nonextensive quantum statistics.

\section*{ACKNOWLEDGMENT}
This work is partly supported by
a Grant-in-Aid for Scientific Research from the Japanese 
Ministry of Education, Culture, Sports, Science and Technology.  

\vspace{0.5cm}
\appendix

\section{The $(q-1)$ EXPANSION IN THE UNNORMALIZED MEM} 
\renewcommand{\theequation}{A\arabic{equation}}
\setcounter{equation}{0}

Tsallis, Sa Barreto and Loh \cite{Tsallis95} developed the AA
to investigate the nonextensivity 
in the observed black-body radiation,
by using the un-normalized MEM \cite{Curado91}.
We will show that the EA with the un-normalized MEM
yields the result in agreement with 
the AA within $O(q-1)$. Calculations of
the $q$-BED and $q$-FDD for $q \simeq 1.0$ are presented.

\subsection{Un-normalized MEM}

An application of the un-normalized MEM to the hamiltonian $\hat{H}$
yields the optimized density matrix given by \cite{Curado91}
\begin{eqnarray}
\hat{\rho}_q &=& \frac{1}{Z_q} [1-(1-q) \beta \hat{H}]^{\frac{1}{1-q}}, \\
Z_q(\beta) &=& Tr\: \{ [1-(1-q) \beta \hat{H}]^{\frac{1}{1-q}} \}.
\label{eq:U21}
\end{eqnarray}
The expectation value of the operator $\hat{O}$ is given by
\begin{eqnarray}
O_q(\beta) &\equiv & \langle \hat{O} \rangle_q 
= Tr \{ \hat{\rho}_q^q \:\hat{O} \}, \\
&=& \frac{1}{Z_q^q}
\:Tr\: \{[1-(1-q) \beta \hat{H}]^{\frac{q}{1-q}}\:\hat{O} \}.
\label{eq:U22}
\end{eqnarray}
 
\subsection{Exact approach}
With the use of the exact representations given by
Eqs. (\ref{eq:B1}) and (\ref{eq:H1}), Eqs. (\ref{eq:U21}) and (\ref{eq:U22}) 
are expressed by
\begin{eqnarray}
Z_q &=& \int_0^{\infty}
G\left(u;\frac{1}{q-1},\frac{1}{(q-1)\beta} \right) Z_1(u) \:du
\hspace{0.5cm}\mbox{for $q > 1$},
\label{eq:U23} \\
&=& \frac{i}{2 \pi} \int_C 
H\left(t; \frac{1}{1-q}, \frac{1}{(1-q)\beta} \right)
Z_1(-t) \:dt
\hspace{0.5cm}\mbox{for $q < 1$},
\\
O_q &=& \frac{1}{Z_q^q} \:\int_0^{\infty}
G\left(u;\frac{1}{q-1}+1,\frac{1}{(q-1)\beta} \right) 
Z_1(u)O_1(u) \:du
\hspace{0.5cm}\mbox{for $q > 1$}, \\
&=&  \frac{i}{2 \pi Z_q^q} \int_C 
H\left(t; \frac{1}{1-q}-1, \frac{1}{(1-q)\beta} \right)
Z_1(-t) O_1(-t) \:dt
\hspace{0.3cm}\mbox{for $q < 1$},
\label{eq:U24}
\end{eqnarray}
with
\begin{eqnarray}
O_1(u) &=& \frac{Tr \{ e^{-u \hat{H} }\:\hat{O} \}}{Z_1(u)}, \\
Z_1(u) &=& Tr \{ e^{-u \hat{H} } \}, 
\end{eqnarray}
where $C$ denotes the Hankel contour,
and $G(u;a,b)$ and $H(t;a,b)$ are given 
by Eqs. (\ref{eq:B7}) and (\ref{eq:H6}), respectively.
In order to evaluate Eqs. (\ref{eq:U23})-(\ref{eq:U24}),
we expand their integrands around $u=\beta$ and $-t=\beta$ 
as is made in Sec. 2.3.
By using Eqs. (\ref{eq:B18}), (\ref{eq:B19}), 
(\ref{eq:T7}) and (\ref{eq:T8}),
we obtain 
\begin{eqnarray}
Z_q &=& Z_1 + \frac{1}{2}(q-1) \beta^2 
\frac{\partial^2 Z_1}{\partial \beta^2}+\cdot\cdot, \\
O_q &=& \frac{1}{Z_q^q} \left[O_1 + (q-1)\beta
\frac{\partial}{\partial \beta}(Z_1 O_1)
+\frac{1}{2}(q-1)\beta^2 \frac{\partial^2 }{\partial \beta^2}(Z_1 O_1)
+\cdot\cdot \right].
\label{eq:V11}
\end{eqnarray}
By using the relations given by
\begin{eqnarray}
\frac{\partial Z_1}{\partial \beta} 
&=& - \langle \hat{H} \rangle_1 Z_1, \nonumber \\
\frac{\partial^2 Z_1}{\partial \beta^2} 
&=& \langle \hat{H}^2 \rangle_1 Z_1, \nonumber \\
\frac{\partial O_1}{\partial \beta} 
&=&  \langle H \rangle_1 \langle O \rangle_1
-\langle \hat{H} \hat{O} \rangle_1, \nonumber \\
\frac{\partial^2 O_1}{\partial \beta^2} 
&=& \langle\hat{H}^2 \hat{O}  \rangle_1 
-  \langle \hat{H}^2 \rangle_1 \langle \hat{O} \rangle_1
+ 2[ \langle \hat{H} \rangle_1^2 \langle \hat{O} \rangle_1
-\langle \hat{H} \hat{O} \rangle_1 \langle \hat{H} \rangle_1], \nonumber
\end{eqnarray}
we finally obtain the $O(q-1)$ expansion of $O_q$ given by
\begin{eqnarray}
O_q &\simeq & O_1 
+(1-q) \left(O_1 \ln Z_1  
+\beta \langle \hat{H} \hat{O} \rangle_1 
+\frac{1}{2}\beta^2 [ \langle \hat{H}^2 \rangle_1 O_1
-\langle \hat{H}^2 \hat{O} \rangle_1] \right)+\cdot\cdot, \nonumber \\
&&
\label{eq:U1}
\end{eqnarray}
which agrees with Eq. (7) of Ref. \cite{Tsallis95} derived by the AA.

\vspace{0.5cm}
\noindent
{\bf (1) $q$-BED}

In order to calculate the $q$-BED, we consider 
$\hat{O}= \hat{n}_k$ with
the hamiltonian for bosons given by
\begin{eqnarray}
\hat{H}= \sum_k (\epsilon_k-\mu) \: \hat{n}_k,
\label{eq:U3}
\end{eqnarray}
where $\hat{n}_k$ and $\epsilon_k$ stand for the number operator and
the energy of the state $k$. We obtain
\begin{eqnarray}
\langle \hat{n}_k \rangle_1 &=& \frac{1}{e^x-1} 
= f_1(\epsilon_k) \equiv f_1,
\hspace{2cm}\mbox{[$x=\beta (\epsilon_k -\mu)$]}
\label{eq:V4} \\
\langle \hat{n}_k \hat{H} \rangle_1 &
=& (\epsilon_k-\mu) e^{x}f_1^2+f_1 E_1, \\
\label{eq:V5} 
\langle \hat{H}^2 \rangle_1 &=& E_1^2 + E_2 + E_3, \\
\label{eq:V6}
\langle \hat{n}_k \hat{H}^2 \rangle_1 &=& 2 (\epsilon_k-\mu)^2 f_1^3 
- 2 (\epsilon_k-\mu) f_1^2 E_1 +f_1 (E_1^2 + E_2 + E_3),
\label{eq:V7}
\end{eqnarray} 
with
\begin{eqnarray}
E_1 &=& \sum_k (\epsilon_k-\mu) f_1, \nonumber \\
E_2 &=& \sum_k (\epsilon_k-\mu)^2 f_1, \nonumber \\ 
E_3 &=& \sum_k (\epsilon_k-\mu)^2 f_1^2. \nonumber
\end{eqnarray}
Substituting Eqs. (\ref{eq:V4})-(\ref{eq:V7}) 
to Eq. (\ref{eq:U1}), we obtain
\begin{eqnarray}
f_q &\simeq& f_1 
+(1-q)\left(f_1 \ln Z_1 +\beta[(\epsilon_k-\mu)e^x f_1^2  + f_1 E_1] \right)  \nonumber \\
&-& \frac{(1-q)\beta^2}{2} \left[ (\epsilon_k-\mu)^2 e^x(e^x+1) f_1^3
+ 2(\epsilon_k-\mu) e^x f_1^2 E_1 \right]+\cdot\cdot. 
\label{eq:V9}
\end{eqnarray}

Tsallis {\it et. al.} \cite{Tsallis95} employed
a one-component boson hamiltonian given by
\begin{eqnarray}
\hat{H}&=& \hbar \omega \:\hat{n} \equiv \epsilon \:\hat{n},
\end{eqnarray}
which yields
\begin{eqnarray}
\langle \hat{n}_k \rangle_1 &=& \frac{1}{e^x-1} 
\equiv f_1, 
\hspace{2cm}\mbox{($x=\beta \epsilon$)}
\label{eq:U31} \\
\langle \hat{n} \hat{H} \rangle_1 &=& \epsilon (e^x+1) f_1 , 
\label{eq:U32}\\
\langle \hat{H}^2 \rangle_1 &=& \epsilon^2 (e^x+1) f_1^2, 
\label{eq:U33}\\
\langle \hat{n} \hat{H}^2 \rangle_1 
&=& \epsilon^2 (e^{2x}+4 e^x+1) f_1^3.
\label{eq:U34}
\end{eqnarray} 
A substitution of Eqs. (\ref{eq:U31})-(\ref{eq:U34}) 
to Eq. (\ref{eq:U1}) leads to
\begin{eqnarray}
f_q &\simeq& f_1 
+(1-q)\left[f_1 \ln Z_1 + x(e^x+1)f_1^2 
- \frac{1}{2} x^2 e^x (e^x+3) f_1^3 \right]+\cdot\cdot. 
\label{eq:U26}
\end{eqnarray}
which is different from Eq. (\ref{eq:V9}) with $\mu=0$
because of the difference in the adopted hamiltonians given by
Eqs. (\ref{eq:U3}) and (\ref{eq:U31}).

\vspace{0.5cm}
\noindent
{\bf (2) $q$-FDD}

We consider $\hat{O}= \hat{n}_k$ with
the hamiltonian for fermions given by 
\begin{eqnarray}
\hat{H}= \sum_k (\epsilon_k-\mu) \: \hat{n}_k,
\label{eq:U27}
\end{eqnarray}
which leads to
\begin{eqnarray}
\langle \hat{n}_k \rangle_1 &=& \frac{1}{e^x+1} 
= f_1(\epsilon_k) \equiv f_1, 
\hspace{2cm}\mbox{[$x=\beta (\epsilon_k -\mu)$]}
\label{eq:U4}\\
\langle \hat{n}_k \hat{H} \rangle_1 &=& (\epsilon_k-\mu) f_1 (1-f_1)
+f_1 E_1, 
\label{eq:U5} \\
\langle \hat{H}^2 \rangle_1 &=& E_1^2 + E_2-E_3, 
\label{eq:U6}\\
\langle \hat{n}_k \hat{H}^2 \rangle_1 &=& (\epsilon_k-\mu)^2 f_1 (1-f_1)(1-2 f_1)
+ 2 (\epsilon_k-\mu) f_1 (1-f_1) E_1 \nonumber \\
&+&f_1 (E_1^2+ E_2-E_3).
\label{eq:U7}
\end{eqnarray} 
Substituting Eqs. (\ref{eq:U4})-(\ref{eq:U7}) 
to Eq. (\ref{eq:U1}), we obtain
\begin{eqnarray}
f_q &\simeq&  f_1
+(1-q)\left(f_1 \ln Z_1 +\beta[(\epsilon_k-\mu) f_1 (1-f_1) + f_1 E_1] \right) \nonumber \\
&-& \frac{(1-q)\beta^2}{2} \left[ (\epsilon_k-\mu)^2 f_1(1-f_1)(1-2f_1)
+ 2(\epsilon_k-\mu) f_1 (1-f_1) E_1\right]+\cdot\cdot. \nonumber \\
&&
\label{eq:U9}
\end{eqnarray}

When assuming a one-component fermion hamiltonian given by
\begin{eqnarray}
\hat{H}= (\epsilon_k-\mu) \hat{n}_k,
\label{eq:U10}
\end{eqnarray}
we obtain
\begin{eqnarray}
\langle \hat{n}_k \rangle_1 &=& \frac{1}{e^x+1} 
\equiv f_1, 
\hspace{2cm}\mbox{[$x=\beta (\epsilon_k -\mu)$]}
\label{eq:U11} \\
\langle \hat{n}_k \hat{H} \rangle_1 &=& (\epsilon_k-\mu) f_1 , 
\label{eq:U12}\\
\langle \hat{H}^2 \rangle_1 &=& (\epsilon_k-\mu)^2 f_1, 
\label{eq:U13}\\
\langle \hat{n}_k \hat{H}^2 \rangle_1 &=& (\epsilon_k-\mu)^2 f_1.
\label{eq:U14}
\end{eqnarray} 
Substituting Eqs. (\ref{eq:U11})-(\ref{eq:U14}) 
to Eq. (\ref{eq:U1}), we obtain
\begin{eqnarray}
f_q &\simeq& f_1 
+(1-q)\left[f_1 \ln Z_1 +\beta(\epsilon-\mu)f_1
- \frac{1}{2}\beta^2 (\epsilon-\mu)^2 
e^{\beta(\epsilon-\mu)} f_1^2\right]+\cdot\cdot. 
\label{eq:U16}
\end{eqnarray}
The difference between Eqs. (\ref{eq:U9}) 
and (\ref{eq:U16}) is due to the
difference in the adopted hamiltonians given 
by Eqs. (\ref{eq:U27}) and (\ref{eq:U10}).
It is noted that the $(q-1)$ expansion of 
$q$-FDD in the FA is given by
\begin{eqnarray}
f_q^{FA} &\simeq& f_1 - \frac{(1-q)}{2}\beta^2 (\epsilon-\mu)^2
e^{\beta(\epsilon-\mu)}f_1^2+\cdot\cdot,
\end{eqnarray}
whose $O(q-1)$ term corresponds to the last term of Eq. (\ref{eq:U16})
derived by the un-normalized MEM.
This is due to the fact that to adopt
the one-component hamiltonian given by Eq. (\ref{eq:U10})
means to use the factorization approximation
from the beginning.

Equation (\ref{eq:V9}) for $q$-BED and Eq. (\ref{eq:U9}) for $q$-FDD  
are expressed in a unified way as
\begin{eqnarray}
f_q &\simeq& f_1 \nonumber \\
&+& (1-q) \left[f_1 \ln Z_1 
+\beta E_1 \{f_1 +(\epsilon-\mu) \frac{\partial f_1}{\partial \epsilon} \}
- \{(\epsilon-\mu)\frac{\partial f_1}{\partial \epsilon}
+ \frac{1}{2}(\epsilon-\mu)^2 \frac{\partial^2 f_1}{\partial \epsilon^2} \}
\right]+\cdot\cdot, \nonumber \\ &&
\label{eq:V10}
\end{eqnarray}
where $f_1=1/(e^{x} \mp 1)$. 
We note that the $O(q-1)$ term of the generalized distribution 
in Eq. (\ref{eq:T5}) derived by the OLM-MEM corresponds 
to the last term in the bracket of Eq. (\ref{eq:V10}).

\section{SUPPLEMENT TO THE INTERPOLATION APPROXIMATION} 
\renewcommand{\theequation}{B\arabic{equation}}
\setcounter{equation}{0}

\subsection{Analytic expressions of $q$-FDD
for $\vert \beta(\epsilon-\mu) \vert \ll 1$}

We may obtain an expression of the $q$-FDD
for small $x$ [$=\beta(\epsilon-\mu)$] with the use of an expansion
for $f_1(\epsilon\beta)$ given by
\begin{eqnarray}
f_1(\epsilon,\beta) 
&=& \frac{1}{2}+ \sum_{n=1}^{\infty} d_{n,1} \:x^n
\hspace{1cm}\mbox{for $ \vert x \vert < 1 $}.
\label{eq:M20} 
\end{eqnarray}
Substituting Eq. (\ref{eq:M20}) to Eqs. (\ref{eq:M1}) and (\ref{eq:M2}),
and employing Eq. (\ref{eq:B1}) and (\ref{eq:H1}),  
we obtain 
\begin{eqnarray}
f_q^{IA}(\epsilon,\beta) 
&=& \frac{1}{2}+ \sum_{n=1}^{\infty} d_{n,q}\:x^n
\hspace{1cm}\mbox{for $\vert x \vert < 1 $},
\label{eq:M15} 
\end{eqnarray}
with
\begin{eqnarray}
d_{n,q} 
&=& d_{n,1} \:\frac{(q-1)^n \:\Gamma(\frac{1}{q-1}+1+n)}
{\Gamma(\frac{1}{q-1}+1)}
\hspace{1cm} \mbox{for $1 < q < 3$},
\label{eq:M13} \\
&=& d_{n,1} \:\frac{(1-q)^n \:\Gamma(\frac{q}{1-q}+1)}
{\Gamma(\frac{q}{1-q}+1-n)}
\hspace{1cm} \mbox{for $0< q < 1$},
\label{eq:M14} 
\\
&=& q \:d_{n,1}
\hspace{4cm} \mbox{for $n=1$},
\label{eq:M16} \\
&=& q (2q-1) \:d_{n,1}
\hspace{3cm} \mbox{for $n=2$}, 
\label{eq:M17}\\
&=& q (2q-1)(3q-2) \:d_{n,1}
\hspace{1cm} \mbox{for $n=3$}, 
\label{eq:M18}
\end{eqnarray}
where $d_{n,1}=(1/n!)\;\partial^n f_1(\epsilon, \beta)/\partial x^n$
at $x=0$: $d_{1,1}=-1/4$, $d_{2,1}=0$, 
$d_{3,1}=1/48$, {\it etc.}.
Equations (\ref{eq:M15})-(\ref{eq:M18}) lead to
\begin{eqnarray}
f_q^{IA}(\epsilon,\beta) &\simeq& \frac{1}{2}-\frac{q}{4}\: x 
+ \frac{q (2q-1)(3q-2)}{48} x^3 +\cdot\cdot 
\hspace{1cm}\mbox{for $\vert x \vert < 1 $}.
\end{eqnarray}

\subsection{Generalized Sommerfeld expansion in the IA}

In the case of $q>1.0$, Eq. (\ref{eq:C4}) yields
\begin{eqnarray}
\frac{\partial f_q(\epsilon)}{\partial \epsilon}
&=& - \int_0^{\infty} 
G\left(u;\frac{q}{q-1},\frac{1}{(q-1)\beta} \right)
\: \frac{(\epsilon-\mu) e^{u(\epsilon-\mu)}}
{[e^{u(\epsilon-\mu)}+1]^2}\:du.
\label{eq:E3}
\end{eqnarray}
Substituting Eq. (\ref{eq:E3}) to Eq. (\ref{eq:E2}) 
and changing the order of integrations
for $\epsilon$ and $u$,
we obtain 
\begin{eqnarray}
c_{n,q} &=& \frac{\beta^n}{n!}
\int_0^{\infty} 
G\left(u;\frac{q}{q-1}, \frac{1}{(q-1)\beta} \right) 
\:u^{- n} \:du 
\int \frac{x^n e^{x}}{(e^x+1)^2}\: dx.
\label{eq:E4}
\end{eqnarray}
At low temperatures, Eq. (\ref{eq:E4}) reduces to
\begin{eqnarray}
c_{n,q} &=& 
\frac{2(1-2^{1-n}) \zeta(n)}{(q-1)^n} \int_0^{\infty}
G\left(u; \frac{q}{q-1},1 \right) u^{-n}\;du, \\
&=& c_{n,1} \; \frac{\Gamma(\frac{1}{q-1}+1-n)}
{(q-1)^n \Gamma(\frac{1}{q-1}+1)}
\hspace{2cm} \mbox{for even $n$}, 
\label{eq:E7} \\
&=& 0
\hspace{6cm} \mbox{for odd $n$}.
\label{eq:E8}
\end{eqnarray}

The ratio of $c_{n,q}/c_{n,1}$ is given by
\begin{eqnarray}
\frac{c_{n,q}}{c_{n,1}} 
&=& \frac{\Gamma(\frac{1}{q-1}+1-n)}
{(q-1)^n \Gamma(\frac{1}{q-1}+1)}
\hspace{1cm}\mbox{for even $n$},
\label{eq:E9} \\
&=& \frac{1}{2-q}
\hspace{4cm}\mbox{for $n=2$}, \\
&=& \frac{1}{(2-q)(3-2q)(4-3q)}
\hspace{1cm}\mbox{for $n=4$}.
\label{eq:E10}  
\end{eqnarray}

In the case of $q < 1.0$, Eqs. (\ref{eq:H4}) and (\ref{eq:E2}) yield 
\begin{eqnarray}
c_{n,q} &=& 
\frac{2(1-2^{1-n}) \zeta(n)}{(1-q)^n} 
\frac{i}{2 \pi}\int_C
H\left(t;\frac{q}{1-q},1 \right) (-t)^{-n} 
\;dt, \\
&=& c_{n,1} \; \frac{\Gamma(\frac{q}{1-q}+1)}
{(1-q)^n \Gamma(\frac{q}{1-q}+1+n)}
\hspace{1cm}\mbox{for even $n$}, 
\label{eq:H7} \\
&=& 0
\hspace{6cm}\mbox{for odd $n$}, 
\label{eq:H8}
\end{eqnarray}
leading to
\begin{eqnarray}
\frac{c_{n,q}}{c_{n,1}} 
&=& \frac{\Gamma(\frac{q}{1-q}+1)}
{(1-q)^n \:\Gamma(\frac{q}{1-q}+1+n)} 
\hspace{1cm}\mbox{for even $n$}, 
\label{eq:H9}\\
&=& \frac{1}{2-q} 
\hspace{4cm}\mbox{for $n=2$}, \\
\label{eq:H10}
&=& \frac{1}{(2-q)(3-2q)(4-3q)}
\hspace{1cm}\mbox{for $n=4$}.  
\label{eq:H11}
\end{eqnarray}
Equation (\ref{eq:H9}) for $q < 1.0$ is the same 
as Eq. (\ref{eq:E9}) for $q>1.0$ 
if we employ the reflection formula of the gamma function:
\begin{eqnarray}
\Gamma(z) \Gamma(1-z) &=& \frac{\pi}{\sin(\pi z)}.\nonumber
\end{eqnarray}

\subsection{The low-temperature phonon specific heat in the IA}

In the case of $q > 1.0$,  Eqs. (\ref{eq:C2}) and (\ref{eq:M1}) yield
\begin{eqnarray}
C_q &\simeq& k_B \beta^2 
\int_0^{\infty} G\left(u;\frac{q}{q-1},1 \right)
\int_0^{\infty}  
\frac{\rho(\omega)(q-1) (\hbar \omega )^2 u \:
e^{(q-1) \beta \hbar \omega u}}
{[e^{(q-1) \beta \hbar \omega u}-1]^2} \:d\omega \:du, \nonumber \\
%
&=& \frac{9 N_a k_B}{(q-1)^4}\left( \frac{T}{\Theta_D}\right)^3
 \int_0^{\infty}
G\left(u;\frac{q}{q-1},1 \right) u^{-4} \:du 
\int_0^{\infty} \frac{x^4 e^x}{(e^x-1)^2} \:dx, \nonumber \\ 
&& \\
&=& \alpha_q \left( \frac{T}{T_D}\right)^3,
\label{eq:G3}
\end{eqnarray}
with
\begin{eqnarray}
\alpha_q &=& \alpha_1 \:
\frac{\Gamma(\frac{1}{q-1}-3)}{(q-1)^4 \:\Gamma(\frac{q}{q-1})}
\hspace{1cm}\mbox{for $1 < q < 3 $ }\label{eq:G4},
\label{eq:G5}
\end{eqnarray}
where $ T_D$ ($= \hbar \omega_D/k_B$) stands for the
Debye temperature and $\alpha_1$ is the $T^3$ coefficient
of the low-temperature specific heat for $q=1.0$.

In the case of $q < 1.0$, a similar analysis with the use of 
Eqs.  (\ref{eq:C2}) and (\ref{eq:M2}) leads to 
\begin{eqnarray}
C_q &\simeq&   k_B \beta^2 \left( \frac{i}{2 \pi} \right)
\int_C H\left(t;\frac{q}{1-q},1 \right) \int_0^{\infty} 
\frac{\rho(\omega)(1-q)(\hbar \omega)^2 (-t) 
e^{-(1-q)\beta \hbar \omega t}}
{[e^{-(1-q)\beta \hbar \omega t}-1]^2} \:d\omega\:dt,\nonumber \\ 
&& \\
&=& \frac{9 N_a k_B}{(1-q)^4}\left( \frac{T}{T_D} \right)^3
\left( \frac{i}{2 \pi}\right) 
\int_C H\left(t;\frac{q}{1-q},1 \right)(-t)^4 \:dt
\int_0^{\infty} \frac{x^4 e^x}{(e^x-1)^2} \:dx,
\end{eqnarray}
from which we obtain
\begin{eqnarray}
\alpha_q &=& \alpha_1 \:
\frac{\Gamma(\frac{q}{1-q}+1)}{(1-q)^4 \Gamma(\frac{q}{1-q}+5)}
\hspace{1cm}\mbox{for $0 < q < 1$}.
\label{eq:G6}
\end{eqnarray}
Equations (\ref{eq:E10}), (\ref{eq:H11}), (\ref{eq:G4}) 
and (\ref{eq:G6}) yield
\begin{eqnarray}
\frac{\alpha_q}{\alpha_1} &=& 
\frac{1}{(2-q)(3-2q)(4-3q)}
= \frac{c_{4,q}}{c_{4,1}}
\hspace{1cm} \mbox{for $0 < q < 4/3$}.
\label{eq:G7}
\end{eqnarray}

\newpage

\begin{table}[htbp]
\begin{center}
\caption{Generalized distributions in the limits of 
$q \rightarrow 1$, $T \rightarrow 0$ and  $\beta \rightarrow 0$}
\vspace{0.5cm}
\renewcommand{\arraystretch}{2.5}
\begin{tabular}{|c|c|c|c|} \hline
method &  $q \rightarrow 1 $ & $T \rightarrow 0$ (FDD) 
&  $\beta \rightarrow 0$ \\ \hline \hline
${\rm EA}^{a}$  & $f_1
+ (q-1)\left[(\epsilon-\mu) \:\frac{\partial f_1}{\partial \epsilon}
+ \frac{1}{2}(\epsilon-\mu)^2 \:\frac{\partial^2 f_1}{\partial \epsilon^2} \right]+\cdot\cdot$  
&  $\Theta(\mu-\epsilon)$  &  $[e_q^{-\beta(\epsilon-\mu)}]^q$ \\ \hline
${\rm IA}^{b}$  & $f_1
+ (q-1)\left[(\epsilon-\mu) \:\frac{\partial f_1}{\partial \epsilon}
+ \frac{1}{2}(\epsilon-\mu)^2 \:\frac{\partial^2 f_1}{\partial \epsilon^2} \right]+\cdot\cdot$  
&  $\Theta(\mu-\epsilon)$  & $[e_q^{-\beta(\epsilon-\mu)}]^q$  \\ \hline
${\rm FA}^{c}$  & $f_1 - \frac{1}{2}(q-1) \beta (\epsilon-\mu)^2
\: \frac{\partial f_1}{\partial \epsilon} +\cdot\cdot$ 
& $\Theta(\mu-\epsilon)$  & $ e_q^{-\beta(\epsilon-\mu)}$ \\ \hline
${\rm SA}^{d}$  & $f_1 + \frac{1}{2} (q-1)(\epsilon-\mu)^2 
\:\frac{\partial^{2} f_1}{\partial \epsilon^2}+\cdot\cdot$  
&  $\Theta(\mu-\epsilon)$  & $ e_q^{-\beta(\epsilon-\mu)}$  \\ \hline
\end{tabular}
\end{center}

\noindent 
$f_1=1/(e^{\beta(\epsilon-\mu)} \mp 1)$: 
$\Theta(x)$, the Heaviside function: 
$e_q^{x}$, $q$-exponential function.

\noindent
$^a$ the exact approach (the present study)

\noindent
$^b$ the interpolation approximation (the present study)

\noindent
$^c$ the factorization approximation \cite{Buy95}

\noindent
$^d$ the superstatiscal approximation \cite{Hasegawa09}
\end{table}

\begin{table}
\begin{center}
\caption{$O(q-1)$ contributions to $c_{n,q}$ ($n=1-4$)
of the generalized Sommerfeld expansion coefficients}
\renewcommand{\arraystretch}{2.5}
\begin{tabular}{|c|c|c|c|c|} \hline

method &  $c_{1,q}$ & $c_{2,q}$ &  $c_{3,q}$ & $c_{4,q}$ \\ \hline \hline
${\rm EA}^{a}$  & 0  
&  $\frac{\pi^2}{6}[1+(q-1)]$ & 0 
& $\frac{7 \pi^4}{360}[1+6(q-1)]$ \\ \hline
${\rm IA}^{b}$  & 0  
&  $\frac{\pi^2}{6}[1+(q-1)]$ & 0  
& $\frac{7 \pi^4}{360}[1+6(q-1)]$ \\ \hline
${\rm FA}^{c}$ &  $\frac{\pi^2}{6}(q-1) $ 
&  $\frac{\pi^2}{6}[1+O((q-1)^2)]$ 
&  $\frac{7 \pi^4}{60} (q-1)$
& $\frac{7 \pi^7}{360}[1+O((q-1)^2)]$ \\ \hline
${\rm SA}^{d}$ &  0  &  $\frac{\pi^2}{6}[1+3 (q-1)]$ 
& 0 & $\frac{7 \pi^4}{360}[1+10(q-1)]$\\ \hline

\end{tabular}
\end{center}

\noindent
$^a$ the exact approach (the present study)

\noindent
$^b$ the interpolation approximation (the present study)

\noindent
$^c$ the factorization approximation \cite{Buy95}

\noindent
$^d$ the superstatiscal approximation \cite{Hasegawa09}

\end{table}

\newpage

\begin{figure}
\begin{center}
\end{center}
\caption{
(Color online)
The temperature dependence 
of $E_q$ of the electron model for $q=1.0$ (dashed curves), 
$q=1.1$ (chain curves), $q=1.2$ (dotted curves)
and $q=1.3$ (solid curves),
the inset showing the enlarged plot for $k_B T/W \leq 0.1$. 
}
\label{fig1}
\end{figure}

\begin{figure}
\begin{center}
\end{center}
\caption{
(Color online)
The $\epsilon$ dependence of the $q$-FDD of
$f_q(\epsilon)$ for $q=0.8$ (solid curves), $q=0.9$ (dotted curves),
$q=1.0$ (dashed curves), $q=1.2$ (double-chain curves),
$q=1.5$ (bold solid curves) and $q=1.8$ (chain curves)
with (a) the linear and (b) logarithmic ordinates,
the results for $q \geq 1.0$ and $q < 1.0$ being calculated 
by the EA and IA, respectively ($k_B T/W=0.1$).
}
\label{fig2}
\end{figure}

\begin{figure}
\begin{center}
\end{center}
\caption{
(Color online)
The $\epsilon$ dependence of the derivative of $q$-FDD,
$-\partial f_q(\epsilon)/\partial \epsilon$, 
for $q=0.8$ (the solid curve), $q=0.9$ (the dotted curve),
$q=1.0$ (the dashed curve), $q=1.2$ (the double-chain curve),
$q=1.5$ (the bold solid curve) and $q=1.8$ (the chain curve)  
with the logarithmic ordinate,
the results for $q \geq 1.0$ and $q < 1.0$ being calculated  
by the EA and IA, respectively ($k_B T/W=0.1$).
}
\label{fig3}
\end{figure}

\begin{figure}
\begin{center}
\end{center}
\caption{
(Color online)
The temperature dependence 
of $E_q$ of the Debye phonon model for $q=1.0$ (dashed curves), 
$q=1.1$ (chain curves), $q=1.2$ (dotted curves)
and $q=1.3$ (solid curves),
the inset showing the enlarged plot for $T/T_D \leq 0.5$. 
}
\label{fig4}
\end{figure}

\begin{figure}
\begin{center}
\end{center}
\caption{
(Color online)
The $\epsilon$ dependence of the $q$-BED of $f_q(\epsilon)$ 
for $q=0.8$ (the solid curve), $q=0.9$ (the dotted curve),
$q=1.0$ (the dashed curve), $q=1.1$ (the chain curve),
$q=1.2$ (the double-chain curve), $q=1.5$ (the bold solid curves) and
$q=1.8$ (the thin solid curve) with the logarithmic ordinate,
the results for $q \geq 1.0$ and $q < 1.0$  being calculated 
by the EA and IA, respectively ($T/T_D=0.01$).
}
\label{fig5}
\end{figure}

\begin{figure}
\begin{center}
\end{center}
\caption{
(Color online)
The $\epsilon$ dependence of the $q$-FDD of $f_q(\epsilon)$ 
calculated by the EA for $q=1.0$ (dashed curves), $q=1.2$ (chain curves),
$q=1.5$ (dotted curves) and $q=1.8$ (solid curves)
with the logarithmic ordinate,
the inset showing the ratio of
$\lambda=f_q^{IA}(\epsilon)/f_q^{EA}(\epsilon)$ ($k_B T/W=1.0$).
}
\label{fig6}
\end{figure}

\begin{figure}
\begin{center}
\end{center}
\caption{
(Color online)
The $\epsilon$ dependence of the $q$-BED of $f_q(\epsilon)$ 
calculated by the EA
for $q=1.0$ (dashed curves), $q=1.1$ (double-chain curves) 
$q=1.2$ (chain curves),
$q=1.5$ (dotted curves) and $q=1.8$ (solid curves)
with the logarithmic ordinate, 
the inset showing the ratio of
$\lambda=f_q^{IA}(\epsilon)/f_q^{EA}(\epsilon)$ ($k_B T/W=0.1$).
}
\label{fig7}
\end{figure}

\begin{figure}
\begin{center}
\end{center}
\caption{
(Color online)
The $\epsilon$ dependence of the $q$-BED of $f_q(\epsilon)$
for $q=1.1$ and 1.2 calculated by the EA (solid curves),
FA (chain curves) and SA (dotted curves)
with the logarithmic ordinate,
$f_1(\epsilon)$ for $q=1.0$ being plotted by the dashed curve
for a comparison ($T/T_D=0.01$).
}
\label{fig8}
\end{figure}

\begin{figure}
\begin{center}
\end{center}
\caption{
(Color online)
The $\epsilon$ dependence of the $q$-FDD of $f_q(\epsilon)$ 
for $q=1.1$ and 1.2 calculated by the EA (the solid curve),
FA (the chain curve) and SA (the dotted curve)
with the logarithmic ordinate, $f_1(\epsilon)$ for $q=1.0$ being 
plotted by the dashed curve for a comparison ($k_B T/W=0.1$).
}
\label{fig9}
\end{figure}

\begin{figure}
\begin{center}
\end{center}
\caption{
(Color online)
The $\epsilon$ dependences of (a) the $q$-FDDs of $f_q(\epsilon)$
and (b) its derivative of $-\partial f_q(\epsilon)/\partial \epsilon$ 
calculated by the IA for $q=0.9$ (solid curves) and 1.1 (bold solid curves),  
and those calculated by the FA for $q=0.9$ (dashed curves) 
and 1.1 (bold dashed curves),
results for $q=1.0$ being plotted by chain curves for a comparison. 
}
\label{fig10}
\end{figure}

\begin{figure}
\begin{center}
\end{center}
\caption{
(Color online)
(a) The $q$ dependence of $c_{n,q}/c_{n,1}$ for $n=2$ and 4 
of the generalized Sommerfeld expansion coefficients
[Eq. (\ref{eq:E6})] with the $q$-FDD, and 
(b) the $q$ dependence of $\alpha_q/\alpha_1$ of the coefficients 
in the low-temperature phonon specific heat with the $q$-BED,
calculated by the EA (circles and squares),
IA (solid curves), FA (dashed curves) \cite{Hasegawa09} and 
SA (chain curves): the result of the SA is 
indistinguishable from that of the FA in (b) (see text).
}
\label{fig11}
\end{figure}

\end{document}